%% file: ms.tex
\documentclass[sigconf]{acmart}

\usepackage{listings}
\usepackage{tikz}
\usepackage{color}
\usepackage{pifont}

\usepackage{graphicx}
\usepackage{pgfplots}
\usepackage{pgfplotstable}

\setcopyright{none}
\renewcommand\footnotetextcopyrightpermission[1]{}

\definecolor{dkgreen}{rgb}{0,0.6,0}
\definecolor{gray}{rgb}{0.5,0.5,0.5}
\definecolor{mauve}{rgb}{0.58,0,0.82}
\definecolor{lightgray}{gray}{0.90}

\begin{document}

\title[The SEED Internet Emulator]{The SEED Internet Emulator and Its Applications in Cybersecurity
Education}

\author{Wenliang Du and Honghao Zeng}
\affiliation{%
  \institution{Syracuse University}
  \city{Syracuse}
  \state{New York}
  \country{USA}
  \postcode{13244}
}
\email{{wedu, hozeng}@syr.edu}

\begin{abstract}
In cybersecurity courses, it is quite challenging to do hands-on activities that involve many
components of the Internet, such as bringing down the Internet, attacking a blockchain, etc. To
solve this problem, we have developed an open-source Internet Emulator, which is a Python
library, consisting of the classes for each essential element of the Internet, including
autonomous system, network, host, router, BGP router, Internet exchange, etc. It also includes
the classes for a variety of services, including Web server, DNS, Botnet, Darknet, Blockchain,
and more are being developed.
Using these classes, users can construct a mini-Internet to emulate the real-world Internet.
Although it is small, it has all the essential elements of the real Internet. The construction
is compiled into Docker container files, and the emulation is executed by Docker on a single
machine, or on multiple cloud machines.

With this Internet Emulator, we can develop a variety of hands-on activities for cybersecurity
courses, including BGP prefix hijacking, attacks on smart contract, using Darknet to achieve
anonymity, launching Botnet and ransomware attacks, etc. While the emulator was initially
developed for cybersecurity courses, it can also be used for network courses, for students to
learn how the Internet technologies work, such as routing, BGP, IP Anycast, and DNS. Many other
interesting network technologies can also be deployed on the emulator, such as content delivery
network and software-defined network.
This is an open-source project: \url{https://github.com/seed-labs/seed-emulator}.
\end{abstract}


\maketitle
\thispagestyle{plain}
\pagestyle{plain}

\input{introduction}

\input{design}

\input{visualization}

\input{evaluation}

\input{application}

\input{related_work}

\section{Summary and Future Work} 

In cybersecurity courses, it is quite challenging to do hands-on activities 
that involve many components of the Internet. To solve this problem,
we have developed an Internet emulator. This paper explains the design of 
the SEED Internet emulator, and presents the evaluation results.  
It also demonstrates how this emulator can be used 
to design hands-on lab activities for Cybersecurity education. 

This is still an ongoing project, and we plan to develop
more components for the emulator, so a wide spectrum
of the Internet services and infrastructures can be included in the emulator. 
We will also improve the visualization tool and develop
more hands-on labs based on the emulator. 
The source code of this open-source project 
is available at GitHub: \url{https://github.com/seed-labs/seed-emulator}.

\begin{acks}
This project was funded in part by the US National Science 
Foundation (No. 1718086) and Syracuse University's 
Meredith Professorship grant and CUSE grant.
\end{acks}

\bibliographystyle{my-ACM-Reference-Format}
\bibliography{bib_emulation}

\end{document}

%% file: introduction.tex
\section{Introduction}

How to design lab activities that involve a large number 
of computers has been a challenging problem in cybersecurity education.
There are two typical solutions to this problem. A common approach
is to use a dedicated infrastructure, such as cyber ranges, consisting
of many physical machines or virtual machines.
NSF has funded some of such infrastructures, such as the DETER project~\cite{deter}.
Many companies and organization also offer cyber range as a service.
These infrastructures, although useful for education,
are quite expensive to build and maintain.
Based on a report~\cite{cyberbit18},
a cyber range costs \$336K to \$576K for the first year,
and \$70K to \$100K for each consecutive year.
Not many institutes can build their own cyber ranges, much less individual
instructors.

Another approach is to use a simulator/emulator to simulate/emulate
the behaviors of a large network and its nodes.
Existing works include
Common Open Research Emulator (CORE)~\cite{CORE2008},
NS-3~\cite{ns3}, GNS-3~\cite{gns3}, Greybox~\cite{greybox},
NetSim (commercial)~\cite{netsim},
OPNET (commercial)~\cite{opnet}, etc.
These simulators/emulators can run on a single or a small number of computers,
making them much more affordable than the infrastructure approach.
This is a much more promising approach.
\textit{We envision that a well-developed Internet emulator, specifically
designed for cybersecurity, will significantly help improve
the hands-on learning in cybersecurity.}

Internet emulation consists of two main challenges: 
the \textit{building} part (how to build the emulator) 
and the \textit{emulation} part (how to conduct the emulation). 
Most of the existing work focuses on the emulation part. This includes
CORE~\cite{CORE2008} (Greybox~\cite{greybox} is based on CORE)
and GNS-3~\cite{gns3}.
They focus primarily on emulating different types of networks (hardware).
While this is necessary for research and product testing,
it is not essential for cybersecurity education,
as most of the cyber attacks are at the higher layers, which
are not specific to network hardware; they are related to 
protocols, services, and applications.

For cybersecurity education, being able to emulate generic network hardware
will be sufficient to cover most scenarios. 
This can be done using the matured Docker and container
technologies~\cite{dockemu}. 
Using these technologies, we can use docker containers to emulate 
hosts and routers, while using virtual networks to emulate the 
networks. These are the building blocks of the Internet. 
Moreover, we can install additional software (such as routing software) 
inside the containers, so the emulated nodes can communicate with one another.
This is the approach taken by Mini-Internet~\cite{mini-internet}.

With the emulation challenge being solved by Docker, 
the primary challenge left for building an Internet emulator
becomes the building part, i.e., how to build an emulation.
This involves building the image for each individual container, 
including specifying how the image should be built,
providing all the needed files for the image, 
configuring the operating system and software 
running inside the image, 
and connecting the container to networks. 
Constructing docker images manually 
is not viable even for a small-size Internet emulator.
It is better to use tools to generate the emulation files. 

The main contribution of our work, also the main 
difference separating our work from the other work,
is how we construct the emulation files. 
We have developed an open-source Python library, consisting of the 
classes for each essential element of the Internet, including
autonomous system, network, host, router, BGP router, Internet exchange, etc. 
It also includes the classes for a variety of services, 
including Web server, DNS, Botnet, Darknet, and Blockchain.
Using these classes, users can construct an Internet 
emulator using Python programs. 
The construction is then compiled into container files 
for Docker to run on a single machine or on multiple machines
from the cloud. 

\textit{Our design is extensible}; 
more components can be added to the emulator.
For example, to run an Ethereum blockchain on the Internet emulator, 
users can manually install and configure the Ethereum system 
inside the emulator (using the low-level APIs provided by our work), but doing so is 
non-trivial. We can develop a new class for this purpose. 
The class will hide all the ``dirty work'' (i.e., the complicated 
configuration part), so deploying an Ethereum blockchain
on the Internet emulator only needs a few lines of code by users.

\textit{Our design is portable}, so a module built in one emulator can be 
independently exported and be reused by another emulator. 
For example, if one has built a DNS infrastructure in
an emulator, he/she can export it. 
Others can plug this DNS infrastructure into their emulators, just like 
plugging a chip into a motherboard. Supporting plugins is an important 
design feature in our work.

\textit{We have built an independent web application} 
to visualize the emulated Internet, as well as allowing users 
to interact with the emulator. 
It can also visualize the events happening inside the emulator. 
For example, in the Morris worm attack (Section~\ref{sec:morris}),
we can visually see how the worm spreads inside the emulated Internet.
In the BGP prefix hijacking attack,
we can visualize the packet flow, so when  
the fake BGP prefixes are announced, we can
clearly see that the packets to the target networks
get redirected.

We call our emulator the SEED Internet Emulator (SEED stands for SEcurity EDucation).  
We will simply call it the emulator in this paper. 
In the rest of this paper, we will discuss the design of our emulator 
and its main elements (Sections~\ref{sec:design-overview} to~\ref{sec:visualization}). 
We have also conducted a comprehensive evaluation 
on the emulator (Section~\ref{sec:evaluation}), 
shedding light on the resource consumption
and the bandwidth that can be achieved by the emulators of different sizes.
To demonstrate the use of the emulator, we show some of the 
interesting hands-on activities that we have 
developed (Section~\ref{sec:applications}).

%% file: design.tex
\section{Overview of the Design} 
\label{sec:design-overview}

In this section, we provide a brief overview of the 
design. Detailed discussions of each element 
in our system are provided in separate sections. 

\subsection{The Programming Approach} 

Among the existing emulators (not necessary
in the field of networking), three approaches are typically 
used to compose an emulation.

\begin{itemize}
  \item GUI approach: use a graphic tool.
  \item Configuration file: describe how the emulator should be built in 
        configuration files, and then use programs to convert 
	them to the emulation files.
  \item Programming: provide libraries and APIs, so users can
        write programs to directly generate the emulation files.
\end{itemize}

These approaches have one thing in common: they all ask users to
use a specific language to build the emulation. They differ 
in their language choices: graphic language, 
configuration language, and programming language. 
The first two types of languages do not require a programming skill,
so they can be used by more people.
Behind the scene, they still have to rely on programming
to produce the final emulation. 
The third approach eliminates the ``middle man'', asking users
to directly write programs. 

The first two approaches lower the barrier to use 
at the cost of expressive power: their expressive power 
cannot match that of a Turing-complete programming language, 
which is the case for most general-purpose
languages, such as C/C++, Java, JavaScript, and Python.
For example, generating 1000 nodes with a slightly different
configuration is quite difficult to achieve using the 
graphic or configuration language, but using programming,
it is just a simple loop.

In our early design, we did use the configuration file approach, but 
as our emulator became more and more complicated, the size of the 
configuration file became too large to manage. Moreover, every time
we needed to make changes to some places, we had to change many
different places. We wished that the language we chose could 
support variables, functions, loops, etc, so the configuration 
could be more concise and managing large configuration could be easier. 
Eventually, we realized that what we wished for 
already exists, and it is called \textit{programming language}.  
Users interested in building Internet emulators are 
very likely those who already have a programming background, 
so programming language is the best choice 
for building such an emulator. We picked the Python
language due to its wide popularity, easiness to use,
and rich set of modules.

\begin{figure}[htb]
  \begin{center}
    \includegraphics[width=0.35\textwidth]{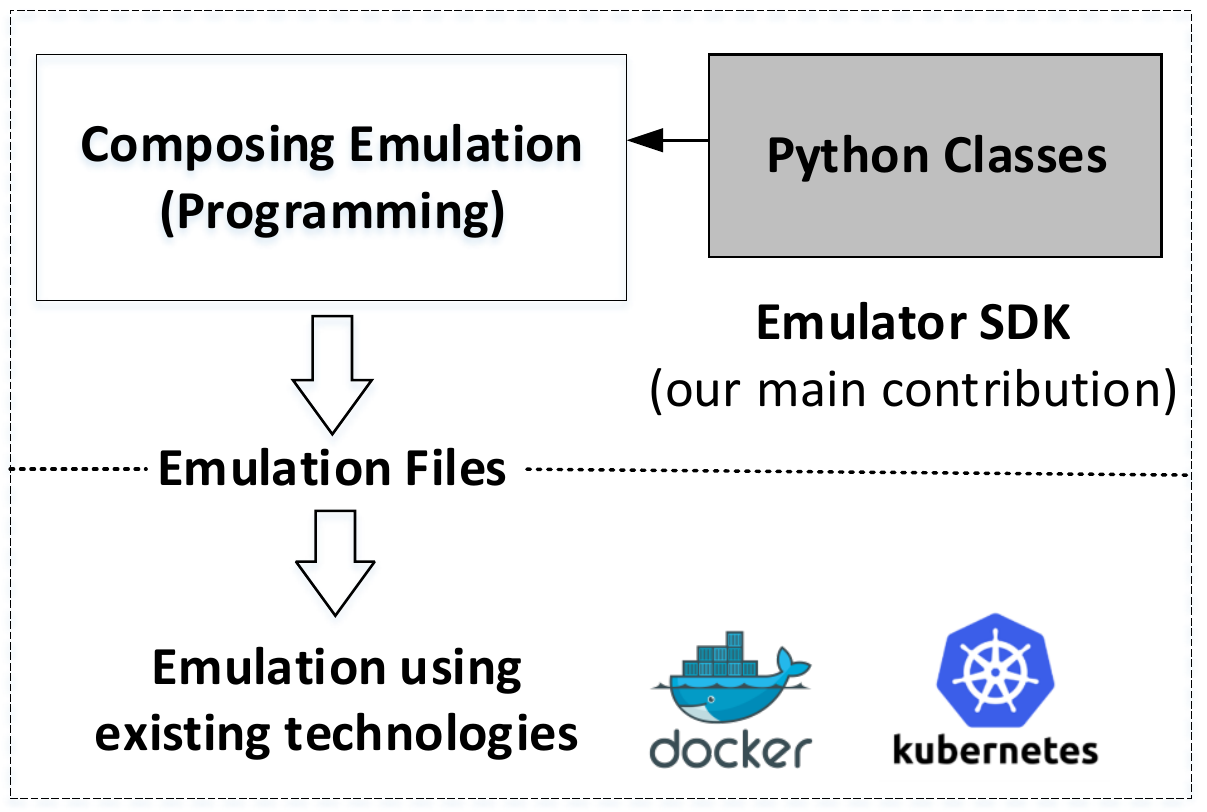}
  \end{center}
  \caption{Our approach}
  \label{fig:overview}
\end{figure}

To help users build an emulator, we need to provide 
basic units to users. For the graphic approach, these 
units are graphic components; for the programming approach, 
they are functions and classes. That is main
focus and contribution of this work: providing 
a rich set of units in the form of Python classes, 
so users can use these units in their programs to
generate the final emulation files. The emulation
can then be conducted by Docker and other technologies. 
Figure~\ref{fig:overview} depicts the overview of our approach.

\subsection{The Architecture} 

\begin{figure}[htb]
  \begin{center}
    \includegraphics[width=0.4\textwidth]{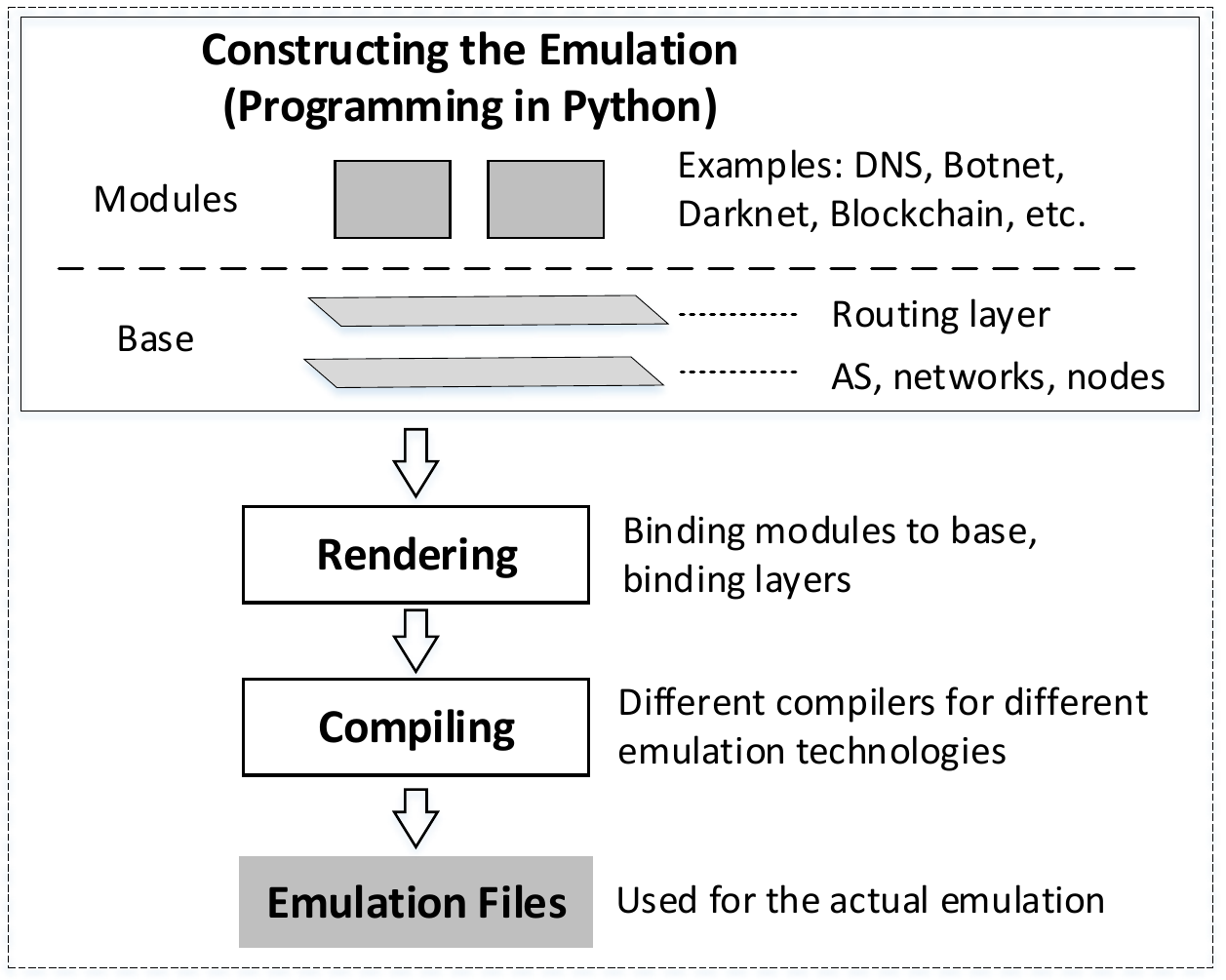}
  \end{center}
  \caption{The architecture}
  \label{fig:architecture}
\end{figure}

The entire emulation construction process can be broken down into 
three phases: composing, binding, and compiling. 
Figure~\ref{fig:architecture} depicts the 
architecture of our design. 

Achieving extensibility and portability are 
important objectives of this project. 
The design should be able to allow new types of 
building units, as well as allowing part of 
an emulator to be reused by another emulator. 
These objectives are achieved through, layers,
plug-ins, and compilers. 

\subsubsection*{The composing phase.}
In this phase, we compose the emulator using the provided 
building units. This is where we create/configure hosts
and networks, connect them, install/configure software
on hosts, set up routing on routers, etc. The 
composition is conducted on layers. The layer idea was inspired by
the image/video editing software such as Adobe Photoshop
and Premier, where each layer contains parts of 
the image/video (in our case, part of the emulation). 

There are two types of layers, base layers and service layers. 
The base layers describe the ``base'' of the topologies, 
like how routers, servers, and networks are connected, how
autonomous systems are peered with each other. 
The service layers describe the high-level services
on the Internet. Examples of services layers are web servers, DNS servers, 
Ethereum blockchain, and botnet. 
No layers are tied to any other layers, meaning each layer can be individually
manipulated, exported, and re-used in another emulation. One can build an entire DNS
infrastructure, complete with root DNS, TLD DNS, and deploy it on any base layer, even with
vastly different underlying topologies.

\subsubsection*{The rendering phase.}
During the composition phase, layers are composed separately, and 
no hard-coded dependency or reference between layers is allowed.
However, layers that install services, like the DNS layer, 
need to know what physical node in the emulation to 
install and configure the service on.
In order to archive this, we introduce the 
\emph{virtual node}, which is merely a string, a symbolic name. 
The layers track the changes using the virtual node name. 
Virtual nodes are not bound to any physical nodes at
the base layer. Decoupling services layers from the base layers
reduces the dependency of layers, making the services layer more portable.
During the rendering phase, all the layers will be merged, and
the virtual nodes will be bound (or deployed) to physical nodes.

\subsubsection*{The compiling phase.}
When composing the Internet emulator, we do not make any assumption
on the underlying emulation technologies. We simply build the 
Internet; how to run the Internet is not the concern of the composition.
That is the compiler's concern. 
The compiler is the one that generates the final emulation files.

Separating the concerns of composition and emulation is important
for an extensible design. 
By implementing different compilers, we can generate the emulation
files for different technologies. 
For this project, we have only implemented three compilers,
one for Docker on a single computer, one for the 
cloud computer (also using Docker), and the other one for 
simply generating the topology diagram. However, 
if we want to run our emulator on other platforms, such 
as ns-3, GNS-3, and CORE, we just 
need to implement a corresponding compiler for them.

\subsection{Visualization} 

For the purpose of education, being able to visualize what is happening
on the Internet is important. For example, to launch the BGP prefix 
hijacking attack, it will be more interesting if students can see
visually how this attack changes the packet flows on the Internet.  
We have develop an independent web-based tool for the visualization
of the emulated Internet, as well as providing a way for users
to interact with the nodes in the emulator.
Users can also record the events occurred on the emulator, and 
replay them in a slower speed, so they can see exactly what has
happened. The tool is presented in Section~\ref{sec:visualization}.

\section{The Base} 
\label{sec:base}

The Internet is formed by host machines, routers, 
networks, Internet exchanges, and autonomous systems.
The autonomous systems peer with one another using
BGP. To build an Internet in an emulation, we need to 
provide the building blocks corresponding to 
these entities.

\subsection{Internet Exchange} 

An autonomous system needs to connect to other autonomous systems,
so they can exchange network traffic. Without the connection,
an AS will be an isolated system.
Through the connections, users from
one autonomous system can reach the users on another AS.
Connecting two autonomous systems is called \textit{peering}.
There are two types of peering, public peering and private peering.
Public peering usually occurs inside a public facility called
Internet Exchange points (IX). Private peering can also be done inside an IX, 
but most private peering occur at colocation center, which is a special type 
of data center.

We emulate both the public and private peering at Internet exchanges. 
An IX is basically a big high-throughput switch that connects the 
routers from different ASes. Through this switch,
packets from one AS can be handed over to another AS. 
In the follow example, we create two Internet exchanges, 
IX-100 and IX-101 (the numbers are the autonomous system 
number assigned to the IXes). Internally, a network 
is created for each IX (their network names are 
automatically set to \texttt{ix100} and \texttt{ix101} 
in our convention. 

\begin{lstlisting}[language=Python, numbers=left]
ix100 = base.createInternetExchange(100)
ix101 = base.createInternetExchange(101)
\end{lstlisting}

\subsection{Stub Autonomous System} 

An autonomous system (AS)
is a collection of connected Internet Protocol (IP) routing prefixes
under the control of one or more network operators on behalf of a single 
administrative entity or domain. 
As an essential element of the Internet, AS is emulated in 
our emulator. 

When composing the emulator, we can create an autonomous system,
and then create hosts, routers, and networks inside it.
In the following example, we create an AS with the 
AS number \texttt{150}; we then create a network,
a router, and two hosts inside the AS. Hosts and 
routers need to be attached to networks. 

\begin{lstlisting}[language=Python, numbers=left]
as150 = base.createAutonomousSystem(150)
as150.createNetwork("net0")
as150.createRouter("router0")
     .joinNetwork("net0").joinNetwork("ix100")
as150.createHost("host0").joinNetwork("net0")
as150.createHost("host1").joinNetwork("net0")
\end{lstlisting}

Routers should be attached to multiple networks. 
In the example above, \texttt{router0} is attached 
to the AS's internal network \texttt{net0} 
and a network called \texttt{ix100}. The second one
is the name of the switch network inside the
Internet exchange \texttt{IX-100}. Therefore,
\texttt{router0} is automatically treated as 
a BGP router in the emulation, and the corresponding
BGP configuration will be conducted on this router.

In the example given above, AS-150 is only connected 
to one Internet exchange, so it does not provide transit
services to others. This is a stub AS. 
It emulates the end customers, such as universities, 
organizations, and most companies.

\subsection{Transit Autonomous System} 

Transit AS is another type of AS connecting to multiple ASes, typically
at multiple locations. It offers to route packets from one AS to another AS,
i.e., it provides transit services to others. 
In our emulator, creating a transit AS is similar to creating a stub
AS. An example is provided in the following.

\begin{lstlisting}[language=Python, numbers=left]
# Create the autonomous system (asn = 2)
as2 = base.createAutonomousSystem(2)

# Create 3 internal networks
as2.createNetwork('net0')
as2.createNetwork('net1')
as2.createNetwork('net2')

# Create 4 routers 
as2.createRouter('r0').joinNetwork('ix100')   (*@\ding{80}@*) 
                      .joinNetwork('net0')
as2.createRouter('r1').joinNetwork('net0')
                      .joinNetwork('ix101')   (*@\ding{80}@*)
                      .joinNetwork('net1')
as2.createRouter('r2').joinNetwork('net1')
                      .joinNetwork('net2')
as2.createRouter('r3').joinNetwork('net2')
                      .joinNetwork('ix102')   (*@\ding{80}@*)
\end{lstlisting}

\begin{figure}[htb]
  \begin{center}
    \includegraphics[width=0.4\textwidth]{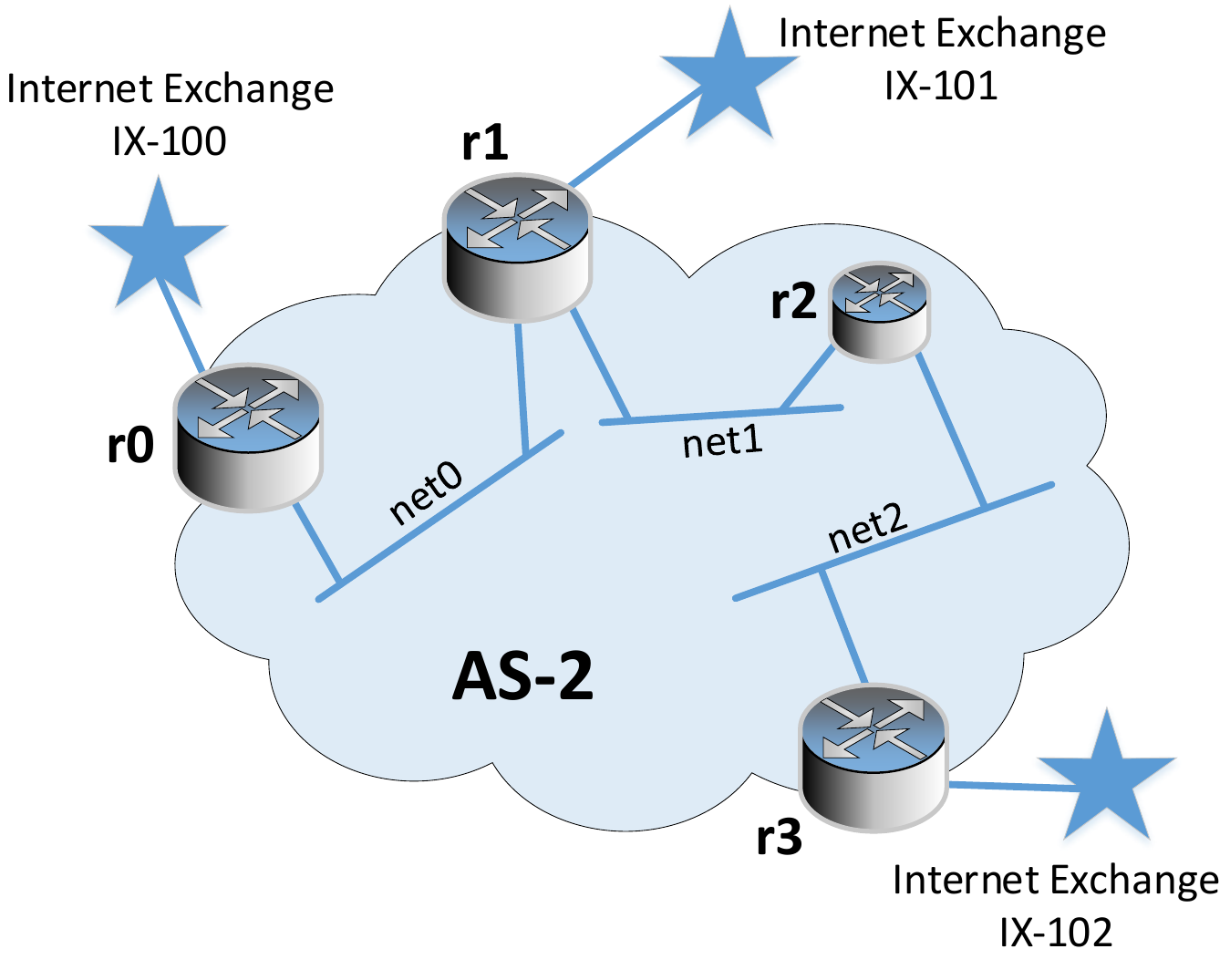}
  \end{center}
  \caption{An example of network topology}
  \label{fig:net-topology}
\end{figure}

In this emulation, we have created three networks, which
are connected using four routers. The network topology is 
depicted in Figure~\ref{fig:net-topology}. Although
this example only shows a simple linear network topology, 
as we can see from the example, more complicated
topologies can also be created using the provided APIs. This is
one of the advantages of using programs to construct an emulation over 
the other approaches, due to the expressive power of a programming language.

The routers \texttt{r0}, \texttt{r1}, and \texttt{r3}
also connect to an Internet exchange network at
IX-100, IX-101, and IX-102, respectively (see the 
lines marked by \ding{80}). 
Therefore, these three routers are BGP routers.
Internally, they will be peered with one another via
the IBGP (Internal BGP) protocol. This is required for 
these BGP routers to exchange the information obtained 
from their respective external peers. 

Inside an autonomous system,
the internal routers talk among themselves, exchanging routing information,
so routers know where to route packets within the AS.
These routers only talk to the routers within the same AS, and
they do not talk to the outsider. Examples of IGP includes OSPF, RIP, 
and IGRP. In our current implementation, we only use the OSPF protocol,
but other protocols can also be supported. 
Therefore, the routers \texttt{r0} to \texttt{r3}
are configured to run the OSPF internal routing protocol.

\subsection{BGP Peering} 

There are two types of BGP peering: EBGP and IBGP peering.
EBGP peering is for BGP routers from different
autonomous systems. 
For the BGP routers in the same
autonomous systems to exchange information, they
also need to peer with each other and run the BGP
protocol to exchange information.
The BGP protocol conducted by the BGP routers
inside the same AS is called Internal BGP (IBGP).
As we have mentioned earlier, in the emulator, 
IBGP peering will be automatically set up. 
All the BGP routers within an AS will peer with one another using IBGP.

For EBGP peering, we need to explicitly specify which ASes 
peer with each other. In the following example, 
AS-3 peers with AS-160, AS-161, and AS-162 at IX-103; 
their peer relationship is provider and customer, with
AS-3 being the providers to the others (i.e., the other 
three ASes are customers of AS-3, and in the real world they
need to pay AS-3 for the 
transit service provided by AS-3). 

\begin{lstlisting}[language=Python, numbers=left]
ebgp.addPrivatePeerings(103, [3],  [160, 161, 162], 
                        PeerRelationship.Provider)
\end{lstlisting}

In a public Internet exchange, autonomous systems want to peer with many other autonomous systems.
Let's say we have N autonomous systems, and they want to peer with one another.
If we use the approach described earlier, each pair of ASes needs
to set up a peering relationship. That will be quite complicated.
Most Internet exchanges provide a mechanism to simplify this. They
provide a special server called \textit{route server}.
All these N autonomous systems will only need to peer with
this route server.
When the router server receives a route from a participant over BGP, it
re-distributes the routes to all other connected participants.
The route server function pretty much like multicast:
any BGP route sent to the router server will
be received by everybody that peers with the route server.

Peering using the route server is supported in the emulator.
In the following example, we peer AS-2, AS-3, and AS-4 at 
IX-100 using the route server approach. One line of code 
will create the peering among these three ASes.

\begin{lstlisting}[language=Python, numbers=left]
ebgp.addRsPeers(100, [2, 3, 4])
\end{lstlisting}

\subsubsection*{Peering relationship and BGP communities.}
What routes a BGP router sends to its peer is based on 
the BGP peering relationship. The emulator supports
two common peering relationship: provider-to-customer
and peer-to-peer. However, to support other 
peering relationship, we have implemented the BGP Large 
communities protocol~\cite{rfc8195} in the emulator. 

This protocol provides a generic solution
for BGP routers to decide which routes should be exported to
or imported from a peer. With this protocol, each
route is marked by labels (community). 
During route exporting and importing, actions can be applied to 
these routes based on which communities they belong to. 
There are many applications of the BGP Large Communities, such as
identifying routes by their geographically locations (countries,
continent etc.), the business relationships between peers (customers,
providers, or peers), and many other aspects. 
In our current implementation,
we only use the BGP Large communities to implement the 
peering relationship, but more interesting applications 
can be supported by our design.

\subsubsection*{The routing software.}
The routing software used in the Emulator is called BIRD,
an acronym standing for ``BIRD Internet Routing Daemon''~\citep{bird}.
BIRD is open source under the GNU General Public License. It
supports a number of standard routing protocols, including
the Border Gateway Protocol (BGP),
the Routing Information Protocol (RIP),
and the Open Shortest Path First protocol (OSPF).
We mainly use BGP and OSPF in the Emulator.

\subsection{Shadow Internet} 

By default, the SEED Internet emulator is isolated from the outside:
machines from outside cannot communicate with those inside the emulator. 
This is done on purpose. In our design, we do provide 
support so users from outside can join the emulation by attaching
their computers to a network inside the emulation. We also provide
the support so packets inside the emulation can be routed out
to the real Internet. 

\begin{figure}[htb]
  \begin{center}
    \includegraphics[width=0.4\textwidth]{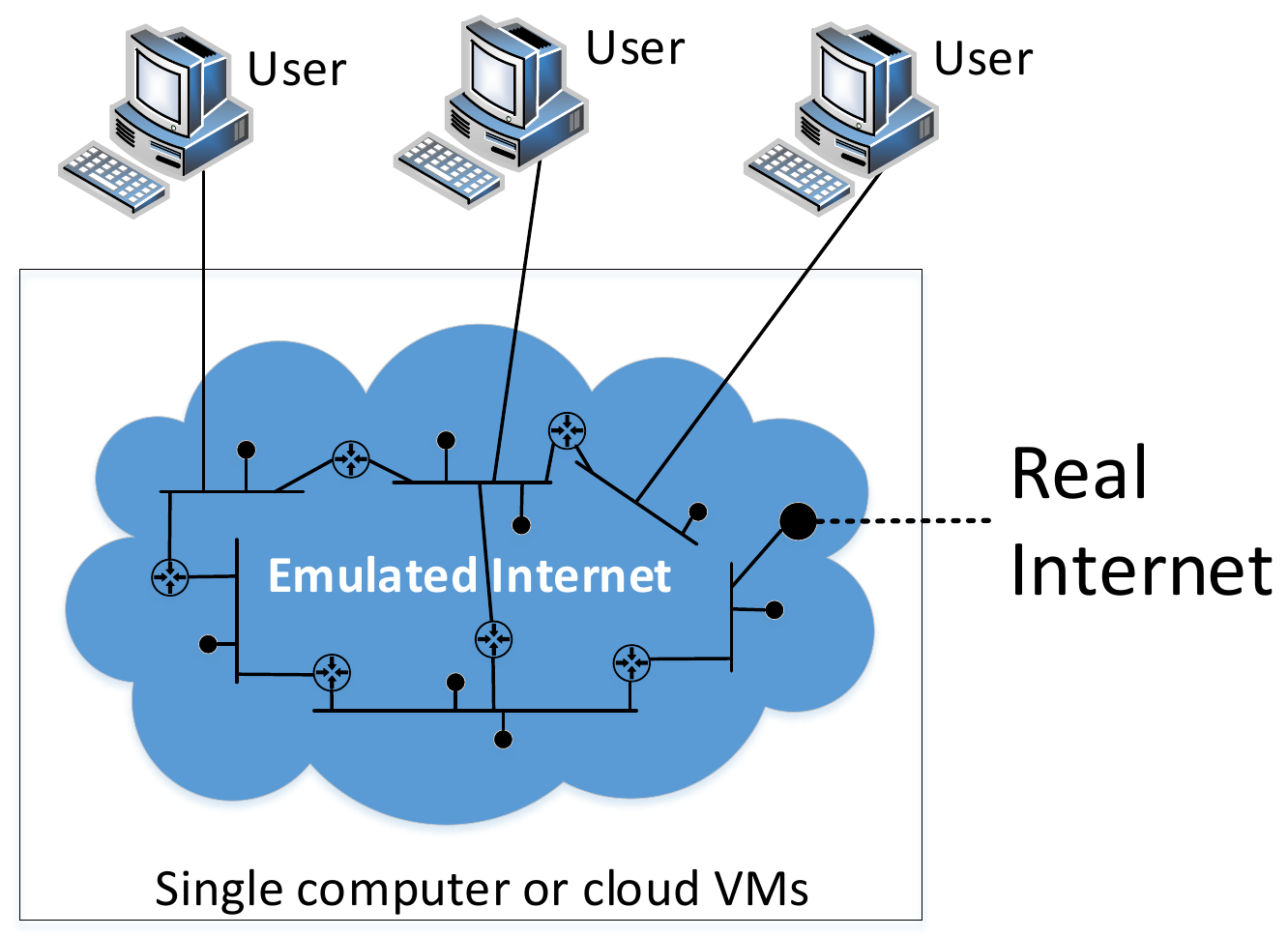}
  \end{center}
  \caption{The shadow Internet}
  \label{fig:shadow-internet}
\end{figure}

When we put them together, the emulator becomes 
a shadow Internet (see Figure~\ref{fig:shadow-internet}). 
Users from outside who want to visit a computer on the real Internet,
such as \texttt{example.com}, will route their \texttt{example.com}-bound
packets into the shadow Internet (by changing the routing table on the user's 
machine). The packets will eventually exit the emulator, into 
the real Internet, and eventually reach its final destination. If we 
want to emulate a real attack against \texttt{example.com}, we can
now do it inside the emulator, which to the user, is equivalent to
the attack inside the real Internet.

To allow outside machines to become a part of the emulator, we just need to
host a VPN server on a node inside the emulator and export the port, 
so it is accessible from the outside. In the following example, we will create a VPN 
server on the \texttt{net0} network inside AS-152. Once an outside
machine VPNs into the network, it essential becomes a host on 
the \texttt{net0} network, and can thus access all the other machines
inside the emulation via the \texttt{net0} network. 

\begin{lstlisting}[language=Python, numbers=left]
as152 = base.getAutonomousSystem(152)
as152.getNetwork('net0').enableRemoteAccess(ovpn)
\end{lstlisting}

To allow the inside machines to reach the real Internet,
we need to create an autonomous system inside the emulator,
so its BGP router can announce the real-world network prefixes
inside the emulator. For example, if we want to allow 
machines inside the emulator to access the real machines
belonging to the Syracuse University, we create 
an autonomous system AS-11827, which is Syracuse University's 
real autonomous system number. We then create a BGP router
for this AS, marking it as \texttt{rw} (real-world), so it will
automatically collect the network prefixes announced by 
AS-11827 in the real world, and announce them inside the emulator. 
We attach this BGP router to IX-102 and assign \texttt{10.102.0.118} 
as its IP address. See the following example: 

\begin{lstlisting}[language=Python, numbers=left]
as11872 = base.createAutonomousSystem(11872)
as11872.createRealWorldRouter('rw').
        joinNetwork('ix102', '10.102.0.118')
\end{lstlisting}
 
Once the above is set up,
packets (from inside the emulator) going to Syracuse University's 
networks will be routed to this AS, and
then be forwarded to the real world (through a NAT server). 
Returning packets will come back from the outside, enter the emulator at
this AS, and be routed to its final destination inside the emulator.

\subsection{APIs on Host} 

To support various applications inside the emulator,
we need to be able to install more software and file
on the hosts, as well as being able to conduct 
configuration. While this can be done during the runtime,
i.e., when the emulator is running, it can be time-consuming 
if we have do these on many hosts. In our design,
we provide APIs so these tasks can be done when
we construct the emulator. The following 
example shows some of the useful APIs. 

\begin{lstlisting}[language=Python, numbers=left]
# Get an instance of the host from AS-151
host0 = as151.getHost('host0')

# Insteall software on the host
host0.addSoftware('telnetd').addSoftware('telnet')

# Import a file to the host 
host0.importFile(hostpath="/home/seed/ddos.py", 
                 nodepath="/tmp/ddos.py")

# Create a file on the host
host0.setFile(content="some content", 
              path="/tmp/file.txt")

# This command is executed when the container is built
host0.addBuildCommand('useradd -m -s /bin/bash seed 
                       && echo "seed:dees" | chpasswd')

# Append a command to the start script       
host0.appendStartCommand('cd /bof && /bof/server &')
\end{lstlisting}

\section{Components} 
\label{sec:components}

Building a good and realistic emulator takes time and efforts,
so it will be desirable if the entire or parts of an emulator
can be shared and used as building blocks for other emulators.
Good emulators/simulators come with this kind of pre-built parts.
For example, Matlab's Simulink has many prebuilt components for simulation.
We call these pre-built emulator (or its part) \textit{component},
and they can be added to an emulator.
For example, we can build a DNS infrastructure that contains
5 root servers, 30 top-level domain servers, and 100 domain nameservers.
This DNS infrastructure can be exported as a component.
Any Internet emulator that needs a DNS infrastructure can
use this component, instead of building one from scratch.

\subsection{The Extensible Design}

\begin{figure}[htb]
  \begin{center}
    \includegraphics[width=0.3\textwidth]{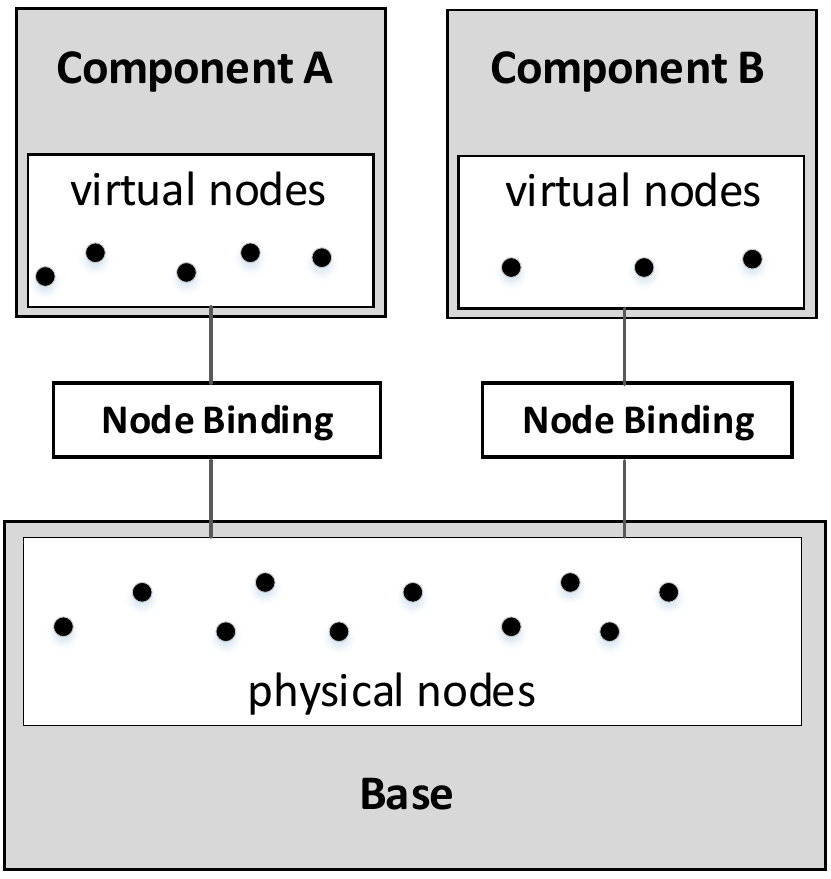}
  \end{center}
  \caption{Making module portable}
  \label{fig:component}
\end{figure}

The biggest challenges in supporting components
are how to make the components portable and
how to connect them to emulators.
First, to support portability, we use dynamic binding
to bind nodes from the service layer to the base layer.
In our layered design, nodes used
at the service layer are virtual nodes, not physical ones.
Physical nodes (containers) are only in the base layer. When we
build each layer, we do not bind them, so each layer
can be built, modified, and exported independently. There
is no dependency among layers. When the emulator construction
is finished, we will render the final emulator. That is
when the layers are merged and
virtual nodes are bound to the physical nodes at
the base layer. See Figure~\ref{fig:component}.

Second, when a component is exported, we would like to allow users to use the
component as a blackbox, without the need to know its internal
logic. To support this, all the virtual nodes of a component
need to be exposed, so
when we connect the component to an emulator, we need to
connect the virtual nodes to the physical nodes.
We draw our inspiration from the circuit design, where complicated components
are enclosed inside a chip, and a chip connects to the rest of the circuit
via pins. We use a similar design pattern, packaging each
component as a ``chip''. Let us use an example to illustrate
the design.

\subsection{Example 1: DNS Infrastructure} 

\begin{figure}[htb]
  \begin{center}
    \includegraphics[width=0.45\textwidth]{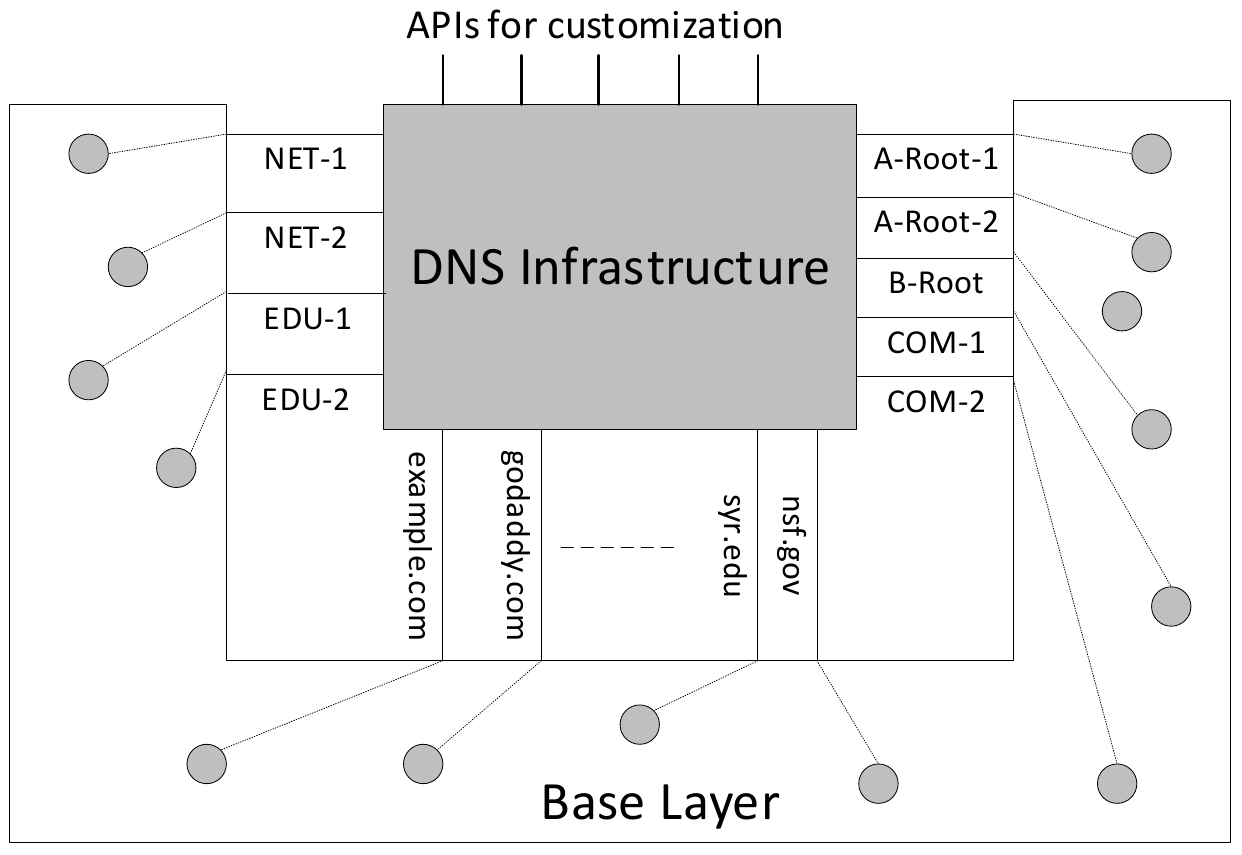}
  \end{center}
  \caption{Examples: DNS component}
  \label{fig:component-dns}
\end{figure}

A DNS component (Figure~\ref{fig:component-dns})
is at the service layer. Its main goal is to set up
the DNS infrastructure, including configuring the zone files for
each domain hosted in this infrastructure. When we build such a
component, we will provide a data sheet to show the purpose
of each pin. When users deploy this DNS component inside an emulator,
they will specify how each pin is connected to
the base layer. For example, they can connect
the \texttt{EDU-1} pin to a randomly selected node inside
the autonomous system 1192 in the base layer, while connecting
the \texttt{nsf.gov} pin to the node with the
IP address \texttt{128.150.4.107}.

In the following, we show how we create an DNS infrastructure.
The \texttt{dns.install(name)} will create a node using 
\texttt{name} if such a node does not exist.
We can install zones on these nodes, and configure their zone files. 
These nodes are virtual nodes; they are just names, 
and they do not bind to any existing physical node in the emulator.
This makes the component portable.

\begin{lstlisting}[language=Python, numbers=left]
# Create a DNS layer
dns = DomainNameService()

# Create two nameservers for the root zone
dns.install('root-a').addZone('.').setMaster()
dns.install('root-b').addZone('.')           
# Create nameservers for TLD zones
dns.install('com-a').addZone('com.').setMaster()
dns.install('com-b').addZone('com.')
dns.install('edu').addZone('edu.')

# Create nameservers for second-level zones
dns.install('ns-example-com').addZone('example.com.')
dns.install('ns-syr-edu').addZone('syr.edu.')
dns.install('ns-google-com').addZone('google.com.')

# Add records to zones
dns.getZone('example.com.').addRecord('@ A 2.2.2.2')
                           .addRecord('www A 5.5.5.5')
                           .addRecord('xyz A 5.5.5.6')
\end{lstlisting}

The above DNS component can be saved into a file and be 
plugged into any base. To do that, we bind each of the 
virtual node in this component to a physical node in the base. 
After that, all the DNS configuration conducted on the virtual node
will be applied to the physical node. The following 
code shows how the binding is conducted. 
For example, the virtual node \texttt{root-a} is bound 
to a host in ASN-171 (the emulator will find a suitable
host for us). After the binding, this selected host
will become a DNS root server. 
 
\begin{lstlisting}[language=Python, numbers=left]
emu.addBinding(Binding('root-a', filter=Filter(asn=171))
emu.addBinding(Binding('root-b', filter=Filter(asn=150))
emu.addBinding(Binding('com-a',  filter=Filter(asn=151))
emu.addBinding(Binding('ns-syr-edu', 
                       filter=Filter(asn=152))
\end{lstlisting}

We have implemented several customized of binders, allowing users
to choose how they want to bind a virtual node. Users can
also use the power of programming language to implement 
any arbitrary type of binding. The following example 
binds the nodes \texttt{bot-node-00}, \ldots, \texttt{bot-node-09} 
to new hosts in randomly selected autonomous systems.   

\begin{lstlisting}[language=Python, numbers=left]
as_list = [150, 151, 152, 153, 154, 160, 161, 162]
for counter in range(10):
  vname = 'bot-node-%.2d'%(counter)
  asn = random.choice(as_list)
  emu.addBinding(Binding(vname, 
      filter=Filter(asn=asn), action=Action.NEW))
\end{lstlisting}

\subsection{Example 2: Blockchain}

We have also implemented an Ethereum-based blockchain component. 
In the following example, we create 6 nodes, with 4 of them
being miners. We also deploy a smart contract from one of the 
nodes. These nodes are virtual nodes. To deploy this blockchain
to an emulator, we just need to bind these virtual nodes
to physical nodes. 

\begin{lstlisting}[language=Python, numbers=left]
# Create Ethereum nodes
e1 = eth.install("eth1").startMiner()
e2 = eth.install("eth2").startMiner()
e3 = eth.install("eth3").startMiner()
e4 = eth.install("eth4").startMiner()
e5 = eth.install("eth5")
e6 = eth.install("eth6")

# Set bootnodes on e1 and e2.
# The other nodes can use these bootnodes to find peers.
e1.setBootNode(True)
e2.setBootNode(True)

# Deploy a smartcontract on e3
contract = SmartContract("./Contracts/contract.bin",
                         "./Contracts/contract.abi")
e3.deploySmartContract(contract)
\end{lstlisting}

One thing that is worthy to mention is that mining is very computational 
intensive, and can quickly consume 100 percent of the CPU,
freezing the entire emulation. To solve this problem, we 
add a \texttt{nice} value to the miner's processes, lowering
their priority. Therefore, even though they still consume 
100 percent of the CPU, they yield to the other 
processes of the emulation.
In the real world, we want the miners
to run as fast as we can (so miners can earn more money),
but in the emulator, we own all the miners, so it does not
matter who ``makes'' money. Lowering the CPU and memory
consumption to allow us to run more miners is more important
in the emulation.

\subsection{Developing a Library of Components}

Apart from the two component examples, we have developed
a few more components, including Darknet and Botnet. 
Many more interesting components can be built for our Internet emulator.
We plan to develop a rich library of components. We also envision
that others may add the components they build to this open-source
and extensible platform. There are three types of components 
that we would like to build:
(1) infrastructure-level service components,
(2) simple service components, and
(3) base-layer components.

\subsubsection*{Infrastructure-level service component.} 
It emulates an infrastructure on the Internet, and it typically
involves a large number of nodes.
Examples of this type of components include
DNS infrastructure, Blockchain, Botnet, Darknet (Tor),   
Public-Key Infrastructure, advertisement network (for web tracking), 
SDN (Software Defined Network),
CDN (Content Delivery Network), etc.

\subsubsection*{Simple service component.} 
It involves a single node or
a small number of nodes. 
For example, a domain registration service is used for people to ``purchase'' domains,
we can build a component for it to emulate domain registrars like GoDaddy.com.
We can implement a mail server using the open-source software, so
users can send emails to each other inside the emulated network. 
We can also develop a number of vulnerable websites and
servers, and package each of them into a component. These components
can be added to an emulator for security-related activities.

\subsubsection*{Base layer component.}
It involves autonomous systems, Internet exchanges,
networks, nodes, and routing. They emulate a portion of the Internet,
such as an organization's network, an Internet backbone (regional or national),
and a collection of autonomous systems and their peering relationships.
Building a fictitious component at the base layer is not difficult;
the challenge is how to build one that accurately emulates
its counterpart in the real world.  For example, to build a component
for the Internet-2 national backbone, we need to know its network topology, where (i.e.,
at what Internet exchange) it peers with other autonomous systems,
how its routes are decided, etc. The more data we can get about this
backbone, the more accurately we can emulate it. We plan to
collect this type of data using the BGP looking glasses deployed
around the world. For missing data, we will use our imaginations.

%% file: visualization.tex
\section{Visualization} 
\label{sec:visualization}

We have also designed a web-based tool called Map
to visualize the emulated Internet, as well as providing a way for users
to interact with the nodes in the emulator. 
This is an independent tool. It interacts with the 
emulation via the APIs provided by the docker daemon. 
This tool is also provided to users in the form as 
an independent container. When this container starts,
the map can be accessed from this 
URL: \url{http://localhost:8080/map.html}.

\begin{figure}[htb]
  \begin{center}
    \includegraphics[width=0.45\textwidth]{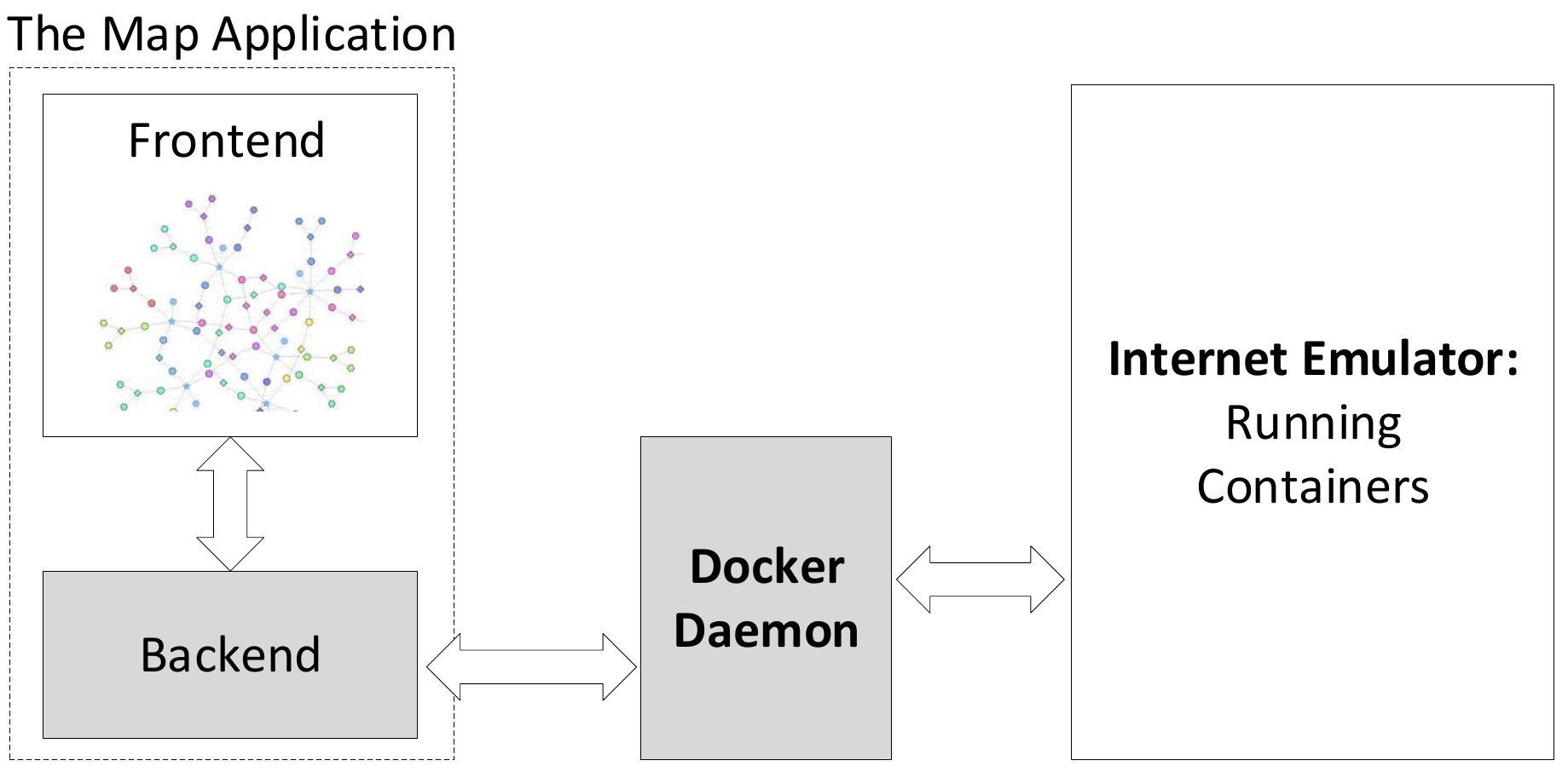}
  \end{center}
  \caption{The architecture of the Map tool}
  \label{fig:map-architecture}
\end{figure}

The architecture of the Map tool is depicted 
in Figure~\ref{fig:map-architecture}. This is a web application
consisting of a front end a backend. The front end is the UI
part running inside a browser, while the backend runs 
on the web server side. The backend retrieves the 
information about the containers and the network
from the Docker daemon (using the APIs provided by Docker),
and send them to the front end for visualization.

\begin{figure*}[t]
  \begin{center}
    \includegraphics[width=0.7\textwidth]{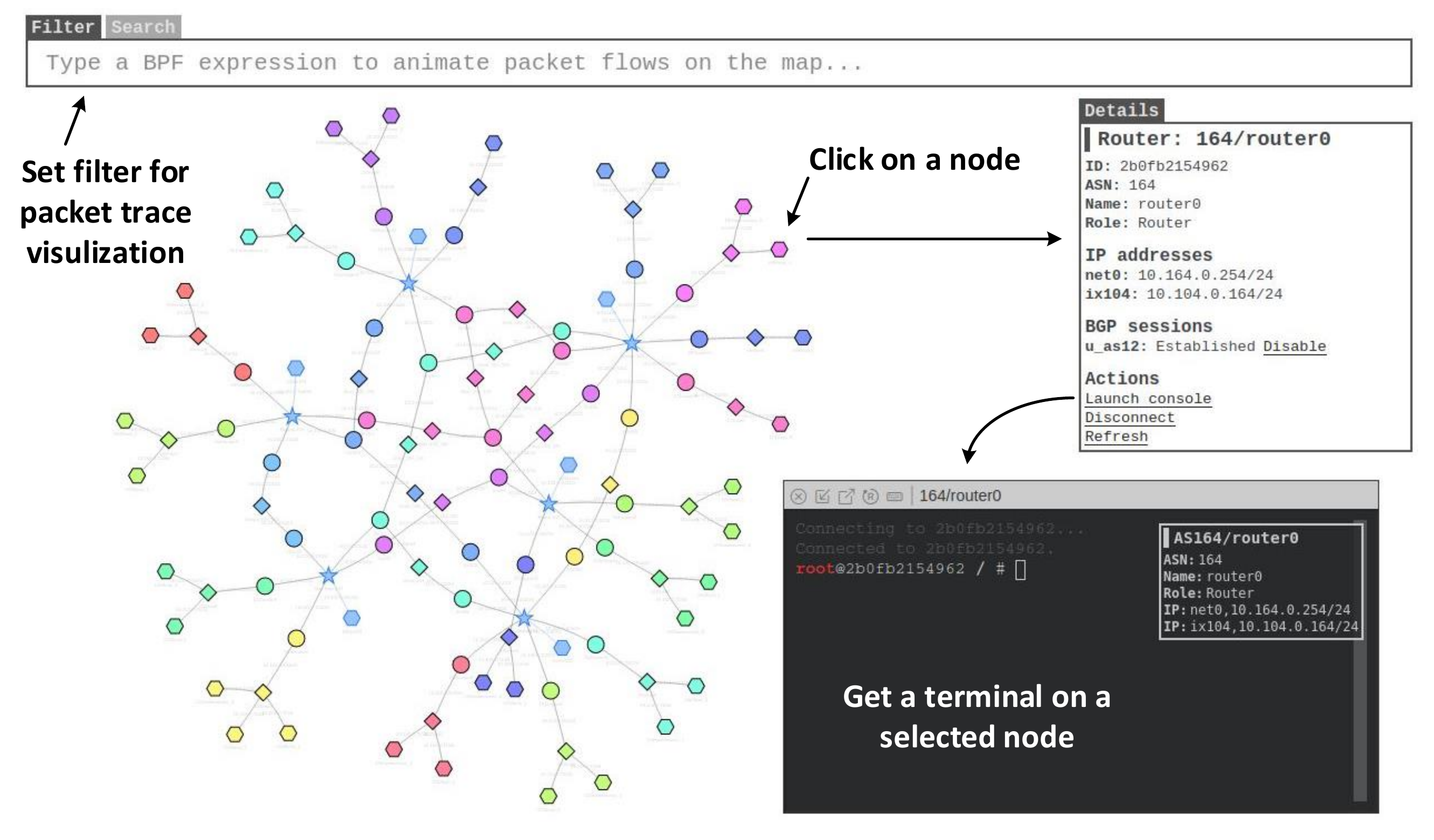}
  \end{center}
  \caption{The visualization tool}
  \label{fig:emulator-gui}
\end{figure*}

When we compile the emulation into final docker files, 
we attach meta data to each container, including
the node’s name, display name, and description, 
list of connected networks, and IP addresses. This 
information can be retrieved by the backend via 
the Docker daemon, and that is how we can visualize
the emulation. 
A picture of the GUI can be found in Figure~\ref{fig:emulator-gui}.

Users can also set filters to visualize network traffic.
The syntax of the filter is the same as that in \texttt{tcpdump}; actually,
the filter is directly fed into the \texttt{tcpdump} program running on all nodes.
When a node sees the packets that satisfy the filter, it
sends an event to the map, which will highlight the node briefly on
the map.

\begin{figure}[htb]
  \begin{center}
        \includegraphics[width=0.49\textwidth]{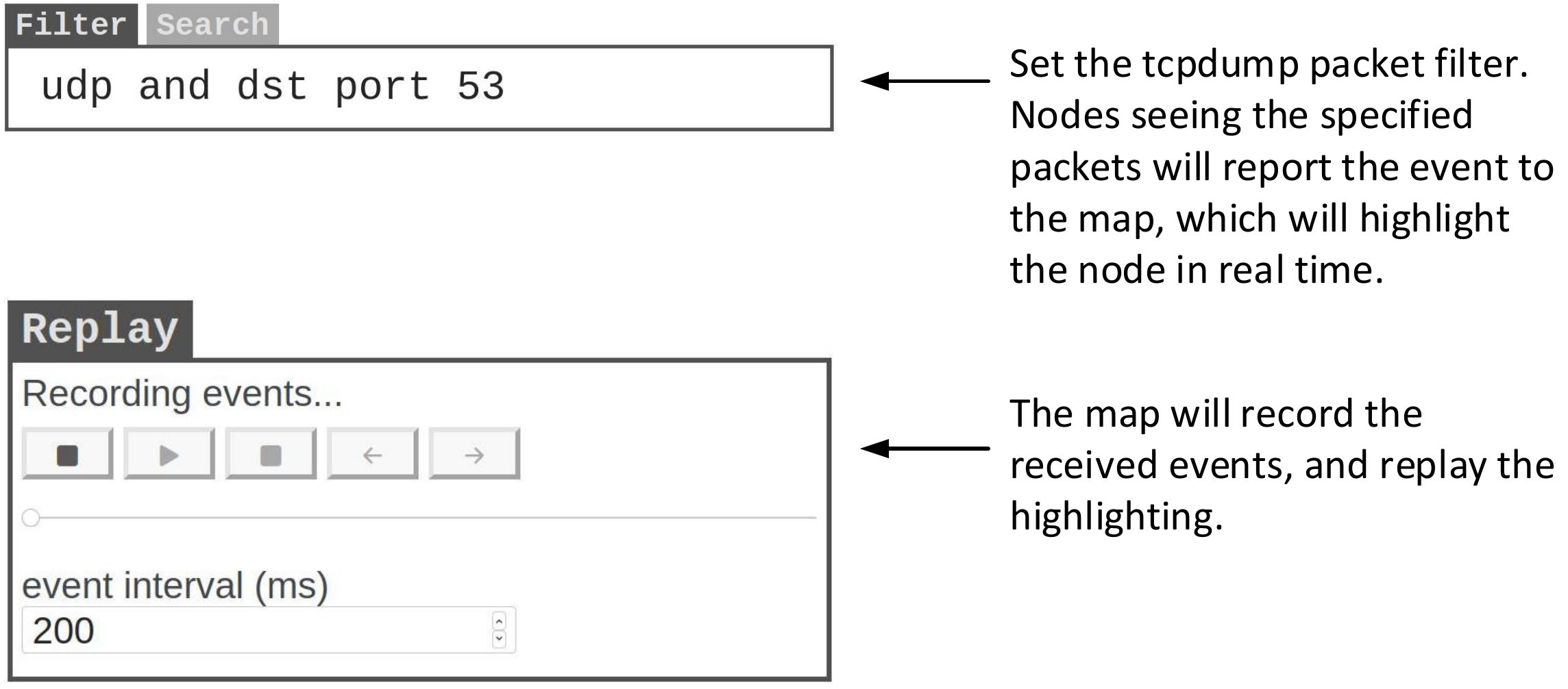}
  \end{center}
  \caption{Capturing and replaying events}
  \label{emulator:fig:replay-event}
\end{figure}

Sometimes, a sequence of events happen too fast to see the actual order
among them. In this case, we can use the Replay panel (see
Figure~\ref{emulator:fig:replay-event}) to record the events and then
replay them at a slower pace. The speed of replaying can be
adjusted by changing the event interval.

\subsubsection*{The extensible design.} 
While capturing the events at the packet level may be sufficient
for many applications, it may not be able to capture the
events at the application level. For example, 
in Blockchain, when a node has successfully mined a block and added 
the block to the blockchain, it is an interesting event that is 
worth visualizing. This event is very difficult to 
capture at the packet level, even though all the information is 
included inside packets. It is better to capture 
the events using the event registration mechanism provided 
by Ethereum. Therefore, to support application-dependent visualization, our design
supports plugins. When a component is developed, a corresponding 
visualization plug-in can also be developed.

%% file: evaluation.tex
\section{Evaluation} 
\label{sec:evaluation}

We have conducted a comprehensive evaluation on the performance of the 
emulator. In this paper, we will only show a subset of the 
evaluation results. Full details of the evaluation, 
can be found in Zeng's Master thesis~\cite{zeng-thesis}.

The performance tests will be conducted in two parts. The first part evaluates the per-unit
resource consumption versus the number of hosts, routers, and networks in emulation. The second
part evaluates the impact on packet forwarding performance when traveling through emulated
nodes. 

Performance information like CPU and memory
usage will be obtained from the \texttt{/proc/stat} and \texttt{/proc/meminfo} files. 
All tests will be conducted without overloading the host system to ensure accurate results. 
All CPU and memory tests result are obtained by first starting the emulation, 
waiting for 60 seconds (to allow route propagation), 
then snapshotting \texttt{/proc/stat} and \texttt{/proc/meminfo} every one second 
for the next 100 seconds. Average CPU utilization and memory are then used as 
the result for each test.
In order to make sure the physical memory is fully utilized, the \texttt{swapness} is set to 1.
Evaluations were conducted in two different environments.

\begin{itemize}
\item On a powerful server: it has two Intel Xeon E5-2660-v2 CPU (40 threads; 20 cores in
  total), with 128 GB of DDR3 memory.

\item On a personal computer: 
  a virtual machine with 2 Intel Xeon Silver 4210 CPU @ 2.00GHz virtual CPUs
  and 4 GB of DDR4 memory. 
\end{itemize}

\subsection{Number of ASes} 
\label{sec:mem-cpu-vs-as}

In this test, there is just one BGP router (no other hosts) in each autonomous system. Each
Internet exchange houses five autonomous systems, which forms full-mesh peering 
within the internet exchange, using the \texttt{Unfiltered} relationship, i.e.,
each AS will peer with the other four. 
Figures~\ref{fig:mem-vs-as} and~\ref{fig:cpu-vs-as} 
show the CPU and memory usage when the number of ASes increases from 0 to 1,000.  

\input{Figs_tex/mem_vs_as.tex}
\input{Figs_tex/cpu_vs_as.tex}

The CPU usage is linear to the number of ASes and it is quite low. Even with 600 ASes, 
the CPU usage is still under $1\%$ (note: our test machine has 20 cores). 
Since the host memory is 128 GB, it reaches the physical limit after the number of ASes
reaches 700. The host system starts to swap memory pages to the disk. That is when the 
CPU usage jumps due to the swapping. 
For the memory consumption, the function used to fit the memory graph 
is $y = 0.1782 * x^2 + 33.06 * x + 10.57$, where $y$ is the memory consumption in MB. 

\begin{itemize}
\item The constant part $10.57$ represents the memory consumes by the host system even when it is not
doing anything.

\item The linear part $33.06 * x$ represents the memory consumption for each added 
  AS due to the addition of new network bridges, BGP routers, etc. 

\item For the quadratic part $0.1782 * x^2$, it is safe to assume that this is the 
  memory consumed by the routing table, because the total number of 
  entries in the routing table does increase quadratically. 
\end{itemize}

\subsection{Number of Hosts} 

This evaluation focuses on the memory consumption of the host nodes running some lightweight
services. The number of autonomous systems is fixed to 10, with each AS having 
one BGP router, one internal network, and \texttt{K} number of hosts, with \texttt{K} 
ranging from 0 to 80 (so the total number of hosts is from 0 to 800). 
Every host node will run a nginx web server and also ping a random router 
in the emulator. Figure~\ref{fig:mem-vs-host} shows the results.

\input{Figs_tex/mem_vs_host.tex}

From the results, one can see that the memory consumption for host nodes is 
linear. This result is quite intuitive, because adding new host nodes 
is merely creating a new container in the emulation
that runs some software. In this case, the software running on the host node operates
individually (\texttt{ping} and \texttt{nginx}), meaning adding a new instance of the software
does not make any difference to the existing instance of the software.

The CPU usage is not plotted, but it is linear to the number of hosts, and it is quite 
low. Even with 800 hosts, the CPU usage is still below 1.5 percent.

\subsection{Number of Internal Routers} 

This test looks at the unit performance when the number of autonomous systems is small, but
each autonomous system has a lot of internal routers and a lot of internal networks. In this
test, there are total ten autonomous systems, with five ASes in each Internet exchange.  
We increase the number of internal routers in each AS from 1 to 90.  

\input{Figs_tex/mem_vs_router.tex}

Figure~\ref{fig:mem-vs-router} shows the results. 
Two fits were attempted on the test results, one using
the results for 0 - 30 routers and the other using the 
results for 30 - 100 routers. One may
observe that the memory consumption starts showing a quadratic-like trend when the number of
routers is small. This is because the total number of routes among all routers
increases quadratically to the number or routers. 

However, when the number of routers become large, the trend starts to
become linear. 
One reason for this is that, as the number of routes from the same
autonomous system increases, the number of memory consumed per route will decrease, 
because the underlying BIRD routing software 
uses an efficient way to store routes, 
where routes with the same route attribute will be
``combined.''~\cite{bird}.

\input{Figs_tex/cpu_vs_router.tex}

Figure~\ref{fig:cpu-vs-router}
shows the relationship between numbers of routers and CPU time.
It follows a quadratic trend. Since the IBGP layer configures full-mesh IBGP
between all reachable internal routers, the number of IBGP sessions pairs in terms of the
number of routers is $n_{paris} = (\frac{n_{router} * (n_{router} - 1)}{2})$. This
leads to the quadratic trend. 
If we use MPLS (Multiprotocol Label Switching)~\cite{rfc3031},
we no longer need to have full mesh iBGP peerings, 
and can thus significantly reduce the CPU and memory usage.
MPLS is supported in the emulator.

\subsection{Network Throughput} 
\label{sec:network-evaluation}

We also measure the network performance of the emulator. We use
\texttt{iperf3} and \texttt{ping} to test the throughput and latency. 
\texttt{iperf3} is a network benchmarking tool that can
be used to measure the throughput of the network. 

In each autonomous system, we create 10 internal networks to form a linear
topology. The client and server are placed at the edge, so there 
are 10 hops between them. The default TCP congestion control algorithm
(\texttt{cubic}) is used on the client and server. 
The \texttt{ipref3} tool is used to test the TCP transmitting (TX) and receiving (RX) 
speed using a single connection between the client and server. 
We have created $x$ number of autonomous systems in the emulator, with
$x$ ranging from 1 to 20, so we have $x$ number of parallel streams.  

\input{Figs_tex/network_vs_stream.tex}

Figure \ref{fig:network-vs-pair} shows the relationship between number of parallel streams and
the throughput. The relationship appears logarithmic. With each additional new pair of client and
server, the throughput increase to some extent, but the increase in throughput drops as the
number of streams go up. One may also notice that the speed stopped increasing at around 11
streams at around 13 Gbps. At the same time, when it reaches 13 Gbps, the CPU usage on all
cores reaches around 100\%. This indicates a hardware performance limit.

In another test, we run \texttt{ping} between the client and server
to measure the round-trip time (RTT). 
We configured \texttt{ping} send one ICMP echo message every millisecond,
with a total of 1,000 ping packets being sent out.
Figure~\ref{fig:rtt-vs-stream} shows the RTT and its relationship with the number of
parallel streams. Even when it reaches 20 streams (or 200 routers), the RTT is still below
1.5ms. The relationship itself appears linear. 

\input{Figs_tex/rtt_vs_stream.tex}

The round-trip time in the emulator is much faster than that in the real Internet,
This is expected because the networks in the emulator are all virtual.
In the future, we will allow users to added delay to the virtual networks,
so we can better emulate the real Internet.

\subsection{Performance on the VM Setting} 

The tests conducted so far are carried out on a powerful server. 
We also envision that many users will choose to run 
a smaller scale emulator on their personal computers, inside 
a virtual machine with limited resource. Therefore, we 
also conducted the tests on a virtual machine with the
following hardware: 

\begin{itemize}
\item 2 virtual CPUs: Intel Xeon Silver 4210 CPU @ 2.00GHz

\item 4 GB of DDR4 memory
\end{itemize}

Using the same the test as the one conducted in Section~\ref{sec:mem-cpu-vs-as},
we have measured memory consumption and CPU usage versus the number 
of ASes (one router and one network in each AS). 
Figure~\ref{fig:mem-vs-as-small} shows the memory consumption.
The CPU usage is still quite low (below $1.5\%$ when $x$ reaches 100), 
so we do not plot the result. 
We have also measured the network throughput using the same test
as the one conducted in Section~\ref{sec:network-evaluation}. 
Figure~\ref{fig:network-vs-stream-small} shows the result. 

\input{Figs_tex/mem_vs_as_small.tex}
\input{Figs_tex/network_vs_stream_small.tex}

\subsection{Insights from the Evaluation} 

In most tests, the bottleneck is at memory since the emulation node uses only a small amount of
CPU compared with the amount of memory they used. Unless one is performing heavy tasks inside
the emulator, like running Ethereum miner, they are most likely to hit the physical memory
limit before hitting the CPU limit.

%% file: Figs_tex/mem_vs_as.tex
\begin{figure}[htb]
\centering
\begin{tikzpicture}
\begin{axis}[xlabel=Number of ASes, ylabel=Memory usage (GB), 
             width=0.45\textwidth, height=0.3\textwidth,
	     legend style={at={(0.01,0.90)}, anchor=west}, 
	     legend cell align={left}]
\addplot[mark=x, only marks] coordinates {
    (0,10.566441)
    (10,10.869575)
    (20,11.184330)
    (30,11.563196)
    (40,12.005314)
    (50,12.539456)
    (60,13.090068)
    (70,13.695919)
    (80,14.286700)
    (90,14.912919)
    (100,15.309082)
    (110,15.996476)
    (120,16.807466)
    (130,17.701900)
    (140,18.578886)
    (150,19.360583)
    (160,20.377525)
    (170,21.415158)
    (180,22.403725)
    (190,23.439255)
    (200,24.421489)
    (210,25.663397)
    (220,26.520257)
    (230,27.661047)
    (240,28.800153)
    (250,29.977751)
    (260,31.296236)
    (270,32.468533)
    (280,33.659876)
    (290,34.835230)
    (300,36.266480)
    (310,37.870891)
    (320,39.237662)
    (330,40.601646)
    (340,41.921509)
    (350,43.306478)
    (400,51.818905)
    (450,61.739351)
    (500,72.249954)
    (600,94.254974)
    (700,122.195800)
    (800,124.195800)
    (900,124.154974)
    (1000,124.118905)
};
\addplot[red, domain=0:1000, samples=100]{(178.2314 * x^2 + 33058.9193 * x + 10566441)/1000000};

\end{axis}
\end{tikzpicture}
\caption{Number of ASes vs. memory consumption} 
\label{fig:mem-vs-as}
\end{figure}

%% file: Figs_tex/cpu_vs_as.tex
\begin{figure}[htb]
\centering
\begin{tikzpicture}
\begin{axis}[axis y line* = left, 
	     xlabel=Number of ASes, xmin = -100, xmax = 1100, 
	     ylabel=CPU time (s), ymin = -100, ymax = 2000, 
             width=0.40\textwidth, height=0.25\textwidth,
	     legend style={at={(0.01,0.90)}, 
	     anchor=west}, legend cell align={left}]

\addplot[mark=x, only marks] coordinates {
    (0,3.6627)
    (10,3.12429)
    (20,4.5581)
    (30,4.31384)
    (40,4.11155)
    (50,5.19825)
    (60,5.28716)
    (70,5.32046)
    (80,5.30699)
    (90,6.26546)
    (100,7.5262)
    (110,7.31166)
    (120,8.11756)
    (130,8.5877)
    (140,9.25580)
    (150,10.22496)
    (160,9.21132)
    (170,10.8235)
    (180,12.2545)
    (190,12.17853)
    (200,12.7840)
    (210,14.15916)
    (220,14.18478)
    (230,14.29501)
    (240,14.18427)
    (250,15.29014)
    (260,15.22659)
    (270,18.9281)
    (280,17.24107)
    (290,19.30165)
    (300,19.31536)
    (310,22.17110)
    (320,22.9366)
    (330,23.29838)
    (340,23.7547)
    (350,24.7798)
    (400,26.5595)
    (450,31.25087)
    (500,40.13310)
    (600,42.6022)
    (700,881.29627)
    (800,861.9463)
    (900,972.3860)
    (1000,1618.252)
};

\addplot[red, domain=0:1000, samples=100]{0.0632 * x + 3};

\end{axis}

\begin{axis}[axis x line = none, xmin = -100, xmax = 1100, 
             axis y line = right, ymin = -2, ymax = 40, 
	     ylabel=CPU usage (\%), 
             width=0.40\textwidth, height=0.25\textwidth]
\end{axis}

\end{tikzpicture}
\caption{Number of ASes vs. CPU time} 
\label{fig:cpu-vs-as}
\end{figure}
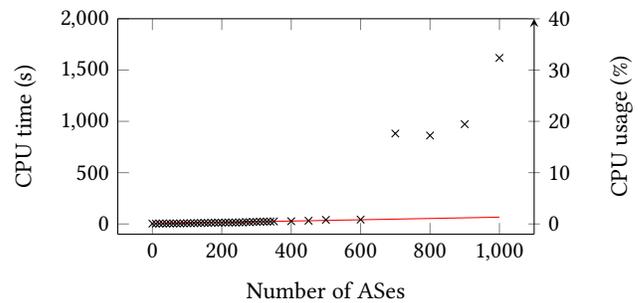

%% file: Figs_tex/mem_vs_host.tex
\begin{figure}[htb]
\centering
\begin{tikzpicture}
\begin{axis}[xlabel=Number of hosts,
             ylabel=Memory usage (GB), 
	     width=0.45\textwidth, height=0.3\textwidth, 
	     legend style={at={(0.01,0.90)}, anchor=west}, legend cell align={left}]

\addplot[mark=x, only marks] coordinates {
    (0,47.389911)
    (50,49.973084)
    (100,52.480038)
    (150,55.426231)
    (200,58.232912)
    (250,61.129955)
    (300,64.263027)
    (350,67.370241)
    (400,70.622884)
    (450,73.907994)
    (500,77.488385)
    (550,80.800987)
    (600,84.455533)
    (650,88.063974)
    (700,91.659013)
    (750,95.371913)
    (800,99.223462)
};

\addplot[red, domain=0:800, samples=80]{(61500 * x + 47389911)/1000000};

\end{axis}
\end{tikzpicture}
\caption{Number of hosts vs. memory consumption} 
\label{fig:mem-vs-host}
\end{figure}
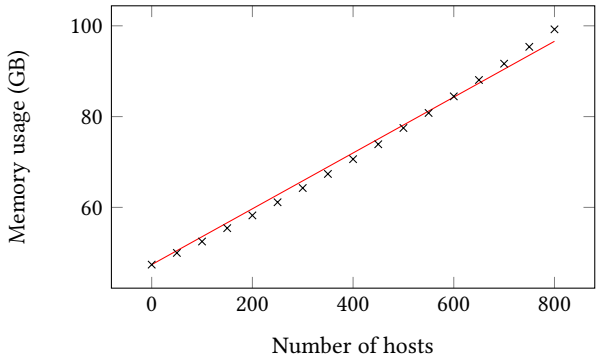

%% file: Figs_tex/mem_vs_router.tex
\begin{figure}[htb]
\centering
\begin{tikzpicture}
\begin{axis}[xlabel=Number of routers, 
             ylabel=Memory usage (GB), 
	     width=0.45\textwidth, height=0.3\textwidth, 
	     legend style={at={(0.01,0.86)}, anchor=west}, legend cell align={left}]

\addplot[mark=x, only marks] coordinates {
    (0,10.730000)
    (10,11.063983)
    (20,11.469105)
    (30,11.878402)
    (40,12.324972)
    (50,12.815662)
    (60,13.333650)
    (70,13.952472)
    (80,14.553774)
    (90,15.212809)
    (100,15.892102)
    (110,16.711962)
    (120,17.508013)
    (130,18.332770)
    (140,19.077838)
    (150,19.979861)
    (160,20.804611)
    (170,21.682796)
    (180,22.607824)
    (190,23.611063)
    (200,24.518727)
    (210,25.511177)
    (220,26.621327)
    (230,27.700441)
    (240,28.847448)
    (250,30.004162)
    (260,31.156020)
    (270,32.338827)
    (280,33.517670)
    (290,34.808775)
    (300,36.083033)
    (400,50.029539)
    (500,63.948325)
    (600,76.262026)
    (700,86.109677)
    (800,96.311972)
    (900,106.473702)
    (1000,113.980153)
};

\addplot[red, domain=0:1000, samples=100]{(154.694893 * x^2 + 38842.83664 * x + 10573163)/1000000};
\addplot[blue, domain=0:1000, samples=100]{(111000*x + 6342004)/1000000};

\end{axis}
\end{tikzpicture}
\caption{Number of internal routers vs. memory usage} \label{fig:mem-vs-router}
\end{figure}
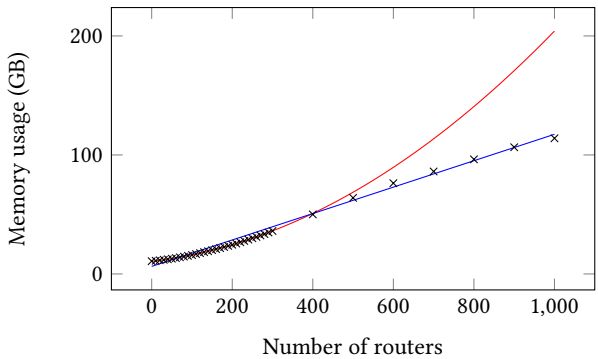

%% file: Figs_tex/cpu_vs_router.tex
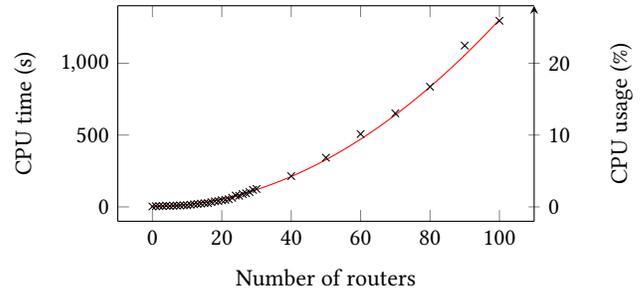
\begin{figure}[htb]
\centering
\begin{tikzpicture}
\begin{axis}[axis y line* = left, ymin = -100, ymax = 1400, xmin = -10, xmax = 110, 
             xlabel=Number of routers, ylabel=CPU time (s), 
	     width=0.4\textwidth, height=0.25\textwidth, 
	     legend style={at={(0.01,0.90)}, anchor=west}, legend cell align={left}]

\addplot[mark=x, only marks] coordinates {
    (0,3.11176)
    (1,5.12763)
    (2,5.15602)
    (3,5.13807)
    (4,8.7357)
    (5,6.31103)
    (6,8.15556)
    (7,9.27287)
    (8,10.28089)
    (9,11.17384)
    (10,11.26598)
    (11,15.16719)
    (12,18.24427)
    (13,19.8042)
    (14,22.8864)
    (15,25.28188)
    (16,26.3141)
    (17,33.16970)
    (18,39.13198)
    (19,40.20674)
    (20,46.29263)
    (21,49.11918)
    (22,53.4550)
    (23,63.2356)
    (24,79.19352)
    (25,77.4129)
    (26,89.29568)
    (27,94.486)
    (28,103.12538)
    (29,118.28847)
    (30,124.950)
    (40,214.12829)
    (50,342.5069)
    (60,507.2210)
    (70,651.29239)
    (80,836.32570)
    (90,1123.16639)
    (100,1295.556)
};

\addplot[red, domain=0:100, samples=100]{0.13 * x^2 + 3.3838};
\end{axis}

\begin{axis}[axis x line = none, axis y line = right, 
             xmin = -10, xmax = 1100, ymin = -2, ymax = 28, 
	     ylabel=CPU usage (\%), 
	     width=0.4\textwidth, height=0.25\textwidth]

\end{axis}
\end{tikzpicture}
\caption{Number of routers per AS vs. CPU time} 
\label{fig:cpu-vs-router}
\end{figure}

%% file: Figs_tex/network_vs_stream.tex
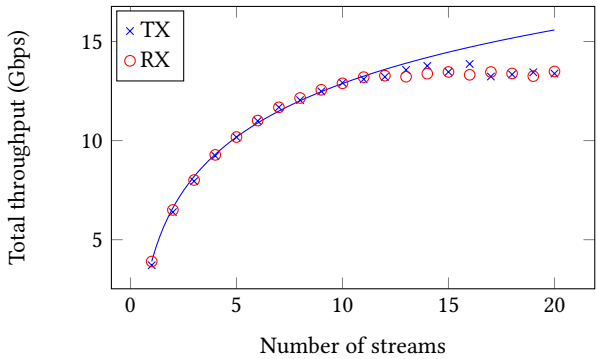
\begin{figure}[htb]
\centering
\begin{tikzpicture}
\begin{axis}[xlabel=Number of streams, 
             ylabel=Total throughput (Gbps), 
	     width=0.45\textwidth, height=0.3\textwidth, 
	     legend style={at={(0.01,0.86)}, anchor=west}, legend cell align={left}]

\addplot[blue, mark=x, only marks] coordinates {
    (1,3.710)
    (2,6.390)
    (3,7.950)
    (4,9.230)
    (5,10.170)
    (6,10.990)
    (7,11.680)
    (8,12.042)
    (9,12.483)
    (10,12.914)
    (11,13.096)
    (12,13.215)
    (13,13.563)
    (14,13.769)
    (15,13.464)
    (16,13.866)
    (17,13.233)
    (18,13.338)
    (19,13.447)
    (20,13.382)
};

\addplot[red, mark=o, only marks] coordinates {
    (1,3.900)
    (2,6.500)
    (3,8.010)
    (4,9.280)
    (5,10.180)
    (6,11.010)
    (7,11.680)
    (8,12.150)
    (9,12.562)
    (10,12.888)
    (11,13.201)
    (12,13.277)
    (13,13.222)
    (14,13.376)
    (15,13.463)
    (16,13.319)
    (17,13.460)
    (18,13.387)
    (19,13.252)
    (20,13.487)
};

\addplot[blue, domain=1:20, samples=100]{(3900 + 3900 * ln(x))/1000};
\legend{TX, RX}
\end{axis}
\end{tikzpicture}
\caption{The transmission and receiving speed} 
\label{fig:network-vs-pair}
\end{figure}

%% file: Figs_tex/rtt_vs_stream.tex
\begin{figure}[htb]
\centering
\begin{tikzpicture}
\begin{axis}[xlabel=Number of streams, 
             ylabel=Average RTT (ms), 
	     width=0.45\textwidth, height=0.3\textwidth, 
	     legend style={at={(0.01,0.90)}, anchor=west}, legend cell align={left}]

\addplot[mark=x, only marks] coordinates {
    (1,0.4110000000)
    (2,0.4785000000)
    (3,0.5056666667)
    (4,0.5580000000)
    (5,0.5646000000)
    (6,0.5863333333)
    (7,0.6085714286)
    (8,0.6206250000)
    (9,0.6432222222)
    (10,0.6654000000)
    (11,0.7200909091)
    (12,0.7685000000)
    (13,0.8685384615)
    (14,0.8863571429)
    (15,0.9263333333)
    (16,1.0074375000)
    (17,1.0587647059)
    (18,1.1177222222)
    (19,1.1588947368)
    (20,1.2308500000)
};

\addplot[red, domain=1:20, samples=100]{0.04088755 * x + 0.33995113};


\end{axis}
\end{tikzpicture}
\caption{The Round-trip time (RTT)} 
\label{fig:rtt-vs-stream}
\end{figure}
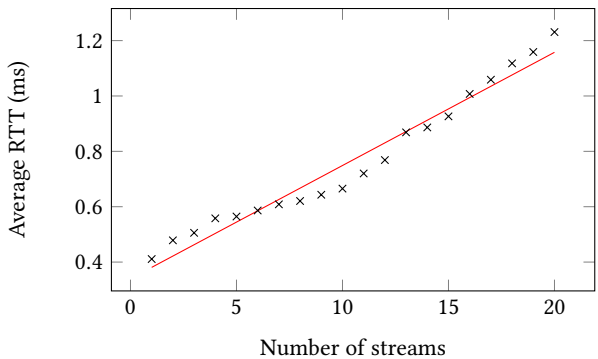

%% file: Figs_tex/mem_vs_as_small.tex
\begin{figure}[htb]
\centering
\begin{tikzpicture}
\begin{axis}[xlabel=Number of ASes, ylabel=Memory usage (GB), 
             width=0.45\textwidth, height=0.3\textwidth, 
	     legend style={at={(0.01,0.90)}, anchor=west}, legend cell align={left}]

\addplot[mark=x, only marks] coordinates {
    (0,0.278342)
    (10,0.874124)
    (20,1.121806)
    (30,1.427956)
    (40,1.958684)
    (50,2.238715)
    (60,2.731775)
    (70,3.161543)
    (80,3.851543)
    (90,3.965252)
    (100,3.849532)
};
\addplot[red, domain=0:100, samples=100]{(178.2314 * x^2 + 33058.9193 * x + 278342)/1000000};

\end{axis}
\end{tikzpicture}
\caption{Memory consumption} 
\label{fig:mem-vs-as-small}
\end{figure}
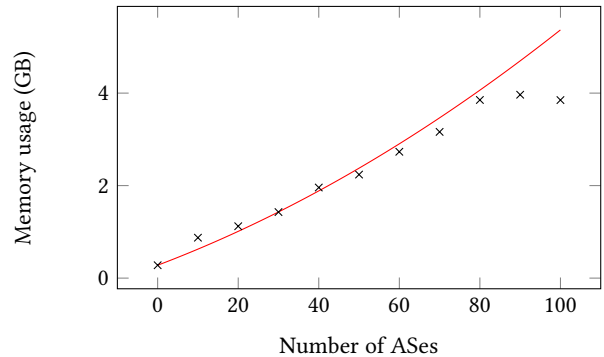

%% file: Figs_tex/network_vs_stream_small.tex
\begin{figure}[htb]
\centering
\begin{tikzpicture}
\begin{axis}[xlabel=Number of streams, 
             ylabel=Total throughput (Gbps), 
	     width=0.45\textwidth, height=0.3\textwidth, 
	     legend style={at={(0.87,0.90)}, anchor=west}, legend cell align={left}]

\addplot[blue, mark=x] coordinates {
    (1,7.050)
    (2,14.291)
    (3,13.696)
    (4,14.280)
    (5,10.616)
    (6,10.964)
    (7,8.495)
    (8,7.354)
    (9,7.354)
    (10,6.003)
};

\addplot[red, mark=o] coordinates {
    (1,7.210)
    (2,14.120)
    (3,13.291)
    (4,13.696)
    (5,11.430)
    (6,9.606)
    (7,7.011)
    (8,5.893)
    (9,4.453)
    (10,3.035)
};
\legend{TX, RX}
\end{axis}
\end{tikzpicture}
\caption{The network throughput} 
\label{fig:network-vs-stream-small}
\end{figure}
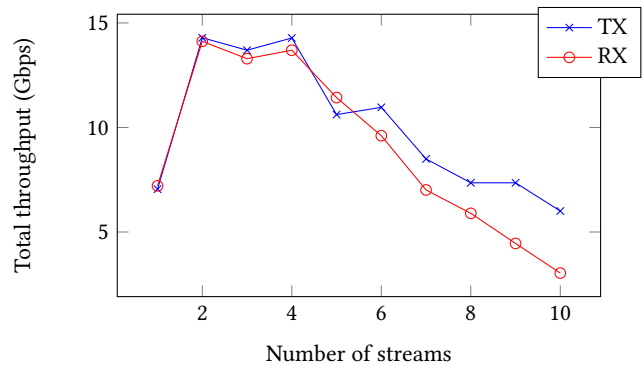

%% file: application.tex
\section{Applications} 
\label{sec:applications}

The Internet emulator can have many applications. Currently,
we mainly focus on the applications on cybersecurity education,
and use it to develop hands-on labs.  
In this section, we give a brief overview on
some of the labs that have already been developed.

\subsection{``Yesterday Once More''} 

Many interesting attacks and incidents happened to the Internet in the past,
and they are very educational. However, as instructors, we are only
able to tell them to students as stories, describing what has happened
technically, on paper. It will be much better if we can 
bring students back to the past to \textit{experience} in person what 
has happened, and to gain the first-hand knowledge. 
This kind of time-travel has been the subject of
many sci-fi movies. Now, we can actually do it using our emulator.

We plan to build emulators to recreate those attacks and incidents,
so students can see with their own eyes the unfolding of the events. They can
observe what has actually happened technically, and
they can even interfere and change the course of the events.
We strongly believe that such a first-hand experience will
significantly enhance students' understanding.
We have so far developed two labs based on the past attacks/incidents, 
but more are being developed. We give a brief overview of 
one of the labs.

\subsection{Morris Worm Attack Lab} 
\label{sec:morris}

\begin{figure}[htb]
  \begin{center}
    \includegraphics[width=0.48\textwidth]{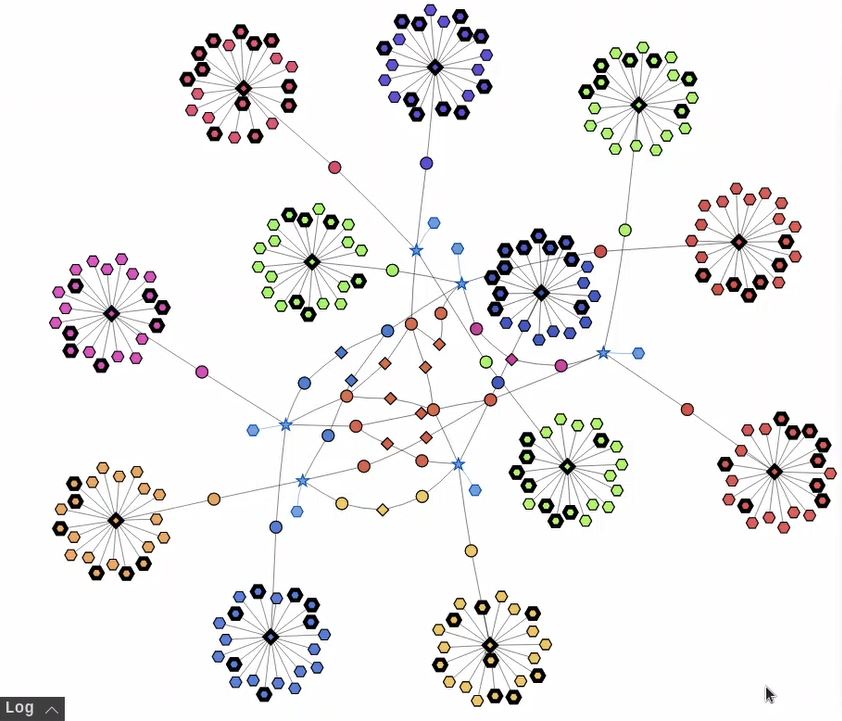}
  \end{center}
  \caption{The spreading of the worm: the nodes with the bold black border are
  compromised nodes, which are also attacking others.}
  \label{fig:morris-worm}
\end{figure}

In this application, we create a mini-Internet emulator using 275 containers, 
consisting of 5 Internet exchanges, 12 stub autonomous systems (AS), 
with each AS having one internal network with 20 hosts (240 hosts in total).
These ASes peer with several transits ASes at the Internet exchanges. 
See Figure~\ref{fig:morris-worm}. During the construction of the emulator,
we installed a vulnerable server on these 240 host containers. 
The server has a buffer-overflow vulnerability. 
The construction only involves 86 lines of Python code.
Despite having these many machines, the entire
emulator can run on a Ubuntu 20.04 virtual machine with 2 cores and 8GB of RAM,
so it is possible to run on most students' personal computers.

The attack code consists of five steps: 
(1) find a  target machine,
(2) launch the buffer-overflow attack on it, 
(3) store a copy of the attack code on the target machine, 
(4) execute the attack code on the target machine, 
(5) go back to the first step and repeat the attack.

We made simplification from the original Morris worm, so the lab can be 
done in a short period of time. However, the main essence of the worm is 
preserved, including the attack, self duplication and propagation. The attack part
directly uses the code from the buffer-overflow lab (one of the popular
SEED labs), so this lab can be used as a follow-up lab, with
the buffer-overflow lab focusing on the attack, and this one 
focusing on the spreading of the worm. 

After compromising a target machine, we also run \texttt{"ping 1.2.3.4"} 
on the target machine. That's how we can visualize which node is compromised
via \texttt{tcpdump}. Figure~\ref{fig:morris-worm} shows the attack
in progress. Here is the reaction from a student after seeing his 
successful attack on the emulator: 
``Seeing it from the entire Internet perspective is remarkable and stunning.''

In our attack code, if we intentionally make the same mistake
as what Morris made, an attack machine can get re-infected repeatedly, each
time, a new instance on the attack will be created, consuming 
additional resources. 
We can observe that once most of the machines are compromised on the emulated
Internet, the CPU usage will become $100\%$. After a while, the host VM
will become slow, and eventually become non-responsive. Basically,
the worm has melted down the emulated Internet. 
That was exactly what the Morris Worm has caused to the real Internet in 1988. 
This lab and its demonstration video
can be found on the SEED website~\cite{morris-lab}.

\subsection{Other Labs} 

We list some of the ideas that we are currently 
investigate. Eventually, we will build a lab
for each of these ideas. 

\begin{itemize}
  \item \textit{The BGP \& Attack Lab.} 
  The goal of this lab is to help students understand how
  BGP ``glues'' the Internet together, and how the Internet is actually
  connected. Students will need to configure some of the BGP
  routers to establish some required peering relationship. 
    Students will also launch the BGP prefix hijacking attacks
    to blackhole a target network. This lab is already published~\cite{bgp-lab}. 

  \item \textit{Blockchain and Smart Contract.}
    Using the blockchain component, we can deploy some vulnerable
    smart contracts in the emulator, and then ask students to 
    launch attacks against them. 

  \item \textit{Botnet Attack.} 
    We have already implemented a prototype of the 
    botnet component, so we can
    easily deploy a botnet~\cite{byob} inside an emulator. We can then
    design lab activities to help students understand how
    this type of attacks works, what it is capable of,
    what techniques it uses to evade detection, etc.

  \item \textit{Darknet.} 
    We have already implemented a prototype of 
    the Darknet component based on
    \texttt{tor}~\cite{tor}. We can deploy it inside an emulator,
    and design lab activities to help students understand how
    the Darknet works, and how it 
    achieves sender's and receiver's anonymity. 

  \item \textit{Firewalls:} we will simulate some realistic
    networks from a fictitious enterprise, and ask students to
    deploy firewalls in the emulation, so these networks
    can be protected. Automated attacks will be built into
    the emulation to test how good the defense is.

\end{itemize}

%% file: related_work.tex
\section{Related Work} 

NS-3 is a discrete-event network simulator, 
targeted primarily for research and educational use~\cite{ns3}. 
We tried to use NS-3 as the base for our work, but 
eventually moved away from it because of a critical
issue: to run programs in NS-3, one needs to use NS-3
versions of the APIs. Since nodes in NS-3 are not real machines, system calls, socket APIs, and
everything that has operating system APIs involved needs to be re-written in NS-3's API to run
inside NS-3. NS-3 attempts to solve this issue by providing the Direct Code Execution (DCE)
framework, which provides emulated Linux APIs that programs can use. However, NS-3
implementations of the APIs are not complete, meaning most of the programs will not be able to
run properly under DCE or require major code changes.

EVE-NG~\cite{eve-ng} and Graphical Network Simulator-3 (GNS-3)~\cite{gns3} 
are two network emulators that use virtual machines to do the emulation.
These two emulators are very similar: they both use a graphical user interface 
to build emulations. Since the emulations are
done by real virtual machines, they can support any kind of nodes, including commercial router
software from vendors like Cisco and Juniper.
While they are widely used by professionals, these emulators do not align well with what 
we want to achieve. One issue is that they both require interacting with
a user interface to build emulation. It is impractical to build large size emulation with a
user interface. Another issue is that 
they focus primarily on emulating different types of networks (hardware).
While this is necessary for research and product testing,
it is not essential for cybersecurity education. Without
this part, Internet emulation can become much simplified. 

Common Open Research Emulator (CORE) is another network emulation tool~\cite{CORE2008}. 
It is also a Python-based emulation framework. 
Its main focus is on running an emulation rather than building emulations.
It also use containers, but it directly uses Linux namespace to create containers, instead
of using Docker. Basically, a significant amount of work implemented in CORE 
can be easily done using Docker. We have chosen to use Docker instead CORE 
as our emulation platform. 

Greybox is a single-host Internet simulator for offline exercise and training networks~\cite{greybox}.
It allows a single host (physical or VM) to provide the illusion of connectivity to the real
Internet: a realistic BGP backbone topology with point-to-point link delays based on the
physical distance between the routers' real-world locations, 
combined with TopGen's application services (HTTP, DNS, email, and more).
Its underlying emulation platform is based on CORE. 

Omur work and Greybox do have many things in common. The
major differences between them are summarized in the following: 
(1) Greybox uses configuration to describe emulations instead of building emulations
programmatically. While configurations is an acceptable format for saving the final 
product of the emulation, it does not offer good portability and reusability.
(2) The way Greybox runs services on hosts is that only one node runs actual services for all
servers in the emulation. The router nodes merely forward traffic to that one single
node. This dramatically reduces the reality of the emulation.
(3) Greybox offers limited options for services.

Mini-internet~\cite{mini-internet} is a project that also targets education. The main goal of the
project is to build a large network with multiple autonomous systems and internet exchanges to
enable students to understand Internet operations alongside their pitfalls.
The project is also built using the container technology. 
The main differences between our work and mini-internet are summarized 
in the following:
(1) The mini-internet project is a good fit for experimenting with the Internet operations, 
but it lacks the ability to host other services, which are essential for cybersecurity education. 
(2) Mini-internet uses configuration to describe emulations, and then 
use programs to convert the configuration to the container files. 
Our work builds the emulation using Python programs. 
In terms of the expressive power, programming languages are much more powerful
than configuration languages.